\documentclass[aps,prd,twocolumn,nofootinbib,superscriptaddress,amsmath,amssymb,floatfix,10pt]{revtex4-1}

\usepackage{graphicx} 
\usepackage{dcolumn} 
\usepackage{bm}  
\usepackage{hyperref} 
\usepackage{adjustbox}
\usepackage{multirow}
\usepackage{comment}
\usepackage{color} 

\usepackage[title]{appendix}

\usepackage[utf8]{inputenc}
\DeclareUnicodeCharacter{223C}{~}
\DeclareUnicodeCharacter{2212}{-}
\DeclareUnicodeCharacter{2009}{ }

\usepackage{gensymb}
\usepackage{siunitx}
\usepackage{physics}
\usepackage{mathtools}

\usepackage{hyperref}
\usepackage{url}

\usepackage[normalem]{ulem}
\usepackage{soul}

\begin{document}
\newcommand{\KMS}{\mbox{km s}^{-1}\,}  
\newcommand{\msol}{\mbox{M}_{\odot}}  
\newcommand{\msun}[0]{{\text{M}_{\sun}}}
\newcommand{\nasa}[0]{NASA}
\newcommand{\nsf}[0]{NSF}
\newcommand{\icecube}[0]{IceCube}
\newcommand{\acron}[0]{ENZO-Einstein}

\def\newacronym#1#2#3{\gdef#1{#3 (#2)\gdef#1{#2}}}

\def\aj{Astronomical Journal}                 % Astronomical Journal
\def\apj{Astrophysical Journal}                % Astrophysical Journal
\def\apjl{Astrophysical Journal Letters}             % Astrophysical Journal, Letters
\def\pasj{PASJ}
\def\apjs{ApJS}              % Astrophysical Journal, Supplement
\def\mnras{MNRAS}            % Monthly Notices of the RAS
\def\prd{Physical Review D}       % Physical Review D
\def\prx{Physical Review X}       % Physical Review X
\def\prl{Physical Review Letters}    % Physical Review Letters
\def\prr{Physical Review Research}    % Physical Review Research
\def\cqg{Classical \& Quantum Gravity}%Classical and Quantum Gravity
\def\nat{Nature}              % Nature
\def\physrep{Physics Reports} % Physics Reports 
\def\na{New Astronomy}		% New Astronomy
\def\aapr{Astronomy \& Astrophysics Reviews}	%Astronomy and Astrophysics Reviews
\def\araa{Annual Reviews of Astronomy \& Astrophysics} 
\def\aap{Astronomy \& Astrophysics}
\def\AstroComp{Astronomy \& Computing}		% Astronomy and Computing 
\def\MNRAS{Monthly Notices of the Royal Astronomical Society}		% Monthly Notices of the Royal Astronomical Society
\def\NatAstro{Nature Astronomy}		% Nature Astronomy 
\def\rmp{Reviews of Modern Physics}		% Reviews of Modern Physics 
\def\sx{SoftwareX}		% SoftwareX 
\def\pasp{Publications of the Astronomical Society of the Pacific} % Publications of the Astronomical Society of the Pacific
\def\joss{Journal of Open Source Software} % Journal of Open Source Software
\def\galax{Galaxies} % Galaxies
\def\rpp{Reports on Progress in Physics} % Reports on Progress in Physics
\def\baas{Bulletin of the American Astronomical Society} % Bulletin of the American Astronomical Society
\def\jpcs{Journal of Physics: Conference Series} % Journal of Physics: Conference Series
\def\ptrs{Philosophical Transactions of the Royal Society} % Philosophical Transactions of the Royal Society 
\def\mpla{Modern Physics Letters A} % Modern Physics Letters A
\def\pdu{Physics of the Dark Universe} % Physics of the Dark Universe
\def\app{Astroparticle Physics} % Astroparticle Physics

\newacronym{\gwsr}{LIGO HIVE}{LIGO High-performance Interferometer Visualization Environment}
%\newacronym{\gwsr}{LOADS}{LIGO Observation and Data Station}
\newacronym{\sdsc}{SDSC}{San Diego Supercomputer Center}
\newacronym{\cra}{CRA}{Center for Relativistic Astrophysics}
\newacronym{\nr}{NR}{numerical relativity}
\newacronym{\ornl}{ORNL}{Oak Ridge National Laboratory}
\newacronym{\lisa}{LISA}{Laser Interferometer Space Antenna}
\newacronym{\ligo}{LIGO}{Laser Interferometer Gravitational Wave Observatory}
\newacronym{\aligo}{aLIGO}{Advanced LIGO}
\newacronym{\lsc}{LSC}{LIGO Scientific Collaboration}
\newacronym{\lvc}{LVC}{LIGO--Virgo Collaboration}
\newacronym{\lvk}{LVK}{LIGO--Virgo--KAGRA Collaboration}
\newacronym{\EM}{EM}{electromagnetic}
\def\lsst#1{Large Synoptic Survey Telescope#1 (LSST#1)\gdef\lsst{LSST}}
\newacronym{\pca}{PCA}{Principal Component Analysis}
\newacronym{\osg}{OSG}{Open Science Grid}
\newacronym{\htc}{HTC}{High-Throughput Computing}

\newacronym{\sph}{SPH}{smooth particle hydrodynamics}
\newacronym{\tsi}{TSI}{Terascale Supernova Initiative}
\newacronym{\wmap}{WMAP}{the Wilkinson Microwave Anisotropy Probe}
\newacronym{\cmbr}{CMBR}{cosmic microwave background}
\newacronym{\ibbh}{IBBH}{intermediate binary black hole}
\newacronym{\hpc}{HPC}{High-performance Computing}
\newacronym{\bssn}{BSSN}{Baumgarte-Shapiro-Shibata-Nakamura}
\def\grb#1{Gamma-Ray Burst#1 (GRB#1)\gdef\grb{GRB}}
\def\sgrb#1{short Gamma-Ray Burst#1 (sGRB#1)\gdef\sgrb{sGRB}}
\def\ns#1{neutron star#1 (NS#1)\gdef\ns{NS}}
\def\frb#1{Fast Radio Burst#1 (FRB#1)\gdef\frb{FRB}}

\def\si#1{Senior Investigator#1 (SI#1)\gdef\si{SI}}
\def\emri#1{Extreme Mass-Ratio Inspiral#1 (EMRI#1)\gdef\emri{EMRI}}
\def\imbh#1{Intermediate Mass Black Hole#1 (IMBH#1)\gdef\imbh{IMBH}}
\def\smbh#1{supermassive black hole#1(SMBH#1)\gdef\smbh{SMBH}}
\def\bbh#1{binary black hole#1 (BBH#1)\gdef\bbh{BBH}}
\def\imbhb#1{intermediate mass black hole binary#1 (IMBHB#1)\gdef\imbhb{IMBHB}}
\def\hmns#1{hypermassive neutron star#1 (HMNS#1)\gdef\hmns{HMNS}}
\def\bh#1{black hole#1 (BH#1)\gdef\bh{BH}}
\def\ns#1{neutron star#1 (NS#1)\gdef\ns{NS}}
\def\hmns#1{hyper-massive neutron star#1 (HMNS#1)\gdef\hmns{HMNS}}
\def\nsbh#1{neutron star-black hole#1 (NSBH#1)\gdef\nsbh{NSBH}}
\def\bns#1{binary neutron star#1 (BNS#1)\gdef\bns{BNS}}
\def\gw#1{Gravitational Wave#1 (GW#1)\gdef\gw{GW}}
\def\fbh#1{final black hole#1 (FBH#1)\gdef\fbh{FBH}}
\def\lhb#1{left-hand boundary#1 (LHB#1)\gdef\lhb{LHB}}
\def\rhb#1{right-hand boundary#1 (RHB#1)\gdef\rhb{RHB}}
\def\HM#1{higher-order modes#1 (HM#1)\gdef\HM{HM}}
\def\pnw#1{post-Newtonian#1 (PN#1)\gdef\pnw{PN}}
\def\eos#1{equation of state#1 (EOS#1)\gdef\eos{EOS}}
\def\gpu#1{graphics processing unit#1 (GPU#1)\gdef\gpu{GPU}}
\def\gr#1{General Relativity#1 (GR#1)\gdef\gr{GR}}
\def\cbc#1{Compact Binary Coalescence#1 (CBC#1)\gdef\cbc{CBC}}

\def\cwb#1{coherent WaveBurst#1 (cWB#1)\gdef\cwb{cWB}}
\def\bcv#1{Bilinear Coupling Veto#1 (BCV#1)\gdef\bcv{BCV}}

\def\dc#1{Detector Characterization#1 (DetChar#1)\gdef\dc{DetChar}}
\def\sei#1{Seismic Isolation subsystem#1 (SEI#1)\gdef\sei{SEI}}

\def\gt#1{Georgia Institute of Technology#1 (GeorgiaTech#1)\gdef\gt{GeorgiaTech}}
\def\umass#1{University of Massachusetts Amherst#1 (UMass#1)\gdef\umass{UMass}}

\newcommand{\code}[1]{\textsl{#1}}

\newcommand{\bam}{\textsl{BAM}}
\newcommand{\whisky}{\textsl{Whisky}}
\newcommand{\carpet}{\textsl{Carpet}}
\newcommand{\cactus}{\textsl{Cactus}}
\newcommand{\kranc}{\textsl{Kranc}}
\newcommand{\maya}{\textsl{Maya}}
\newcommand{\enzo}{\textsl{Enzo}}
\newcommand{\enzop}{\textsl{Enzo-P}}
\newcommand{\cello}{\textsl{Cello}}
\newcommand{\yt}{\textsl{yt}}
\newcommand{\Yt}{\textsl{Yt}}
\newcommand{\spec}{\textsl{SpEC}}

\newcommand{\bw}{{BayesWave}}
\newcommand{\gwcompare}{{gw-compare}}
\newcommand{\rift}{{RIFT}}
\newcommand{\pycbc}{{PyCBC}}
\newcommand{\pygrb}{{PyGRB}}
\newcommand{\lalsuite}{{\it lalsuite}}
\newcommand{\gstlal}{{GstLAL}}

\newcommand{\np}{\vspace{0.1cm}\noindent}

\newcommand{\red}[1]{{\color{red}{#1}}}

\newcommand{\chipave}{\langle \chi_p \rangle}

\newcommand{\GaTech}{\affiliation{School of Physics, Georgia Institute of Technology, Atlanta, Georgia 30332, USA}}
\author{Chad Henshaw}\GaTech
\author{Alice Heranval}\GaTech
\author{Laura Cadonati}\GaTech
\title{Time-frequency structure in the post-merger binary black hole gravitational wave signal}
\date{\today}

\begin{abstract}
Gravitational wave signals from asymmetric binary black hole systems have been shown to exhibit additional chirps beyond the primary merger chirp in the post-merger region of the time-frequency domain. These secondary post-merger chirps correlate to the evolving geometry of the common horizon that forms as the binary merges and were previously studied through numerical relativity simulation in a zero-spin regime. In this work, we investigate the post-merger time-frequency structure in systems with both aligned and precessing spin using widely available waveform models. We find that the inclusion of strong aligned spin $\left(\xi = 0.75\right)$ induces further post-merger time-frequency peaks. Additionally we show that even mild precessing spin $\left(\chi_p = 0.25\right)$ strongly affects the distribution of post-merger radiative power across the celestial sky of the final black hole. Our results support the theory of a correlation between the post-merger signal and horizon geometry.
\end{abstract}

\maketitle

\section{Introduction}

\gw{s} are ripples in spacetime that propagate outward from powerful astrophysical sources, like the merger of black holes, and encode information about their originating systems. The LIGO-Virgo-KAGRA collaboration (LVK) have detected more than 90 gravitational wave signals from \cbc{s} during their first three observing runs (O1-O3)~\cite{gwtc1,gwtc2, gwtc2.1, gwtc3}. These include two \bns{} mergers~\cite{GW170817, GW190425}, three \nsbh{} mergers~\cite{GW190814, NSBHx2}, and additional detections during the fourth observing run (O4), including another potential candidate \nsbh{}~\cite{O4_NSBH}. The remaining events are consistent with \bbh{} mergers, classified primarily by mass estimation using Bayesian inference.\par

The gravitational waveforms emitted by quasi-circular BBH mergers~\cite{Lincoln1990, PhysRevD.52.821, Racine_2008} are characterized by 15 parameters: eight intrinsic (masses $m_1, m_2$ and Cartesian spin components $S_{1\,x,y,z}, S_{2\,x,y,z}$) and seven extrinsic (luminosity distance $d_L$, right ascension $\theta$ declination $\delta$, coalescence time $t_c$, inclination $\iota$, azimuthal angle $\phi$, and polarization $\psi$). The estimation of these parameters via Bayesian inference utilizes waveform models developed from \nr{} simulations and compiled into template banks that form the basis for {\it approximants}. 
Beyond these 15 primary parameters, \gw{} signals can contain additional information whose extraction is mostly limited by the detectors' sensitivity; for instance, the detection of tidal effects and \bns{} post-merger oscillations would constrain the interior equation of state of neutron stars~\cite{Read2009, Blanchet2014,Clark2016, Bauswein2016}.\par 

Black holes, governed by the ``no-hair" theorem, are uniquely characterized by mass, spin, and electrical charge within classical \gr{}, with their interior causally disconnected from the exterior universe beyond the event horizon. Even so, \bbh{} mergers can reveal additional dynamics of the system. 
For instance, the orbital plane precession from misaligned spins can be characterized by $\chi_p$ for single misalignment~\cite{Schmidt_2015} or $\langle \chi_p \rangle$, $\sqrt{\langle \chi_p^2 \rangle}$ for dual misalignment \cite{Gerosa_2021, Henshaw2022, Gerosa2023}, which offers insights into black hole formation channels~\cite{Antonini_2020, Trani_2021, Belczynski_2020}.\par 

In this work, we focus on how the post-merger-horizon geometry manifests itself in the gravitational radiation during the ringdown phase, as the initially deformed common horizon of the \fbh{} relaxes to an axisymmetric state. Numerical simulations have shown that asymmetric systems with mass ratio $q>1$ $\left(q \equiv m_1 / m_2, \: m_1 \geq m_2\right)$ exhibit a distinctive double-chirp pattern in time-frequency domain when viewed from high inclination angles; this behavior connects the horizon geometry to observable GW signals~\cite{CalderonBustillo2020}.\par

The double-chirp pattern is currently only observed in simulations that include \HM{} beyond $(l,m) = (2, \pm 2)$, and it becomes more prominent with increasing total mass $M = m_1 + m_2$ and inclination angle, from $\iota = 0$ (face-on) to $\iota = \pi /2$ (edge-on). Recent search efforts using \HM{}-inclusive waveforms have demonstrated significant improvements in detection efficiency, particularly for massive binaries, although it remains challenging to distinguish these signals from detector noise transients~\cite{Harry2018, Chandra2022,Sharma2022}.\par

Numerical relativity is the key to understanding the relation between the double-chirp pattern and the horizon geometry. \nr{} simulations characterize the dynamical black hole horizon by identifying marginally trapped surfaces~\cite{Penrose1965} - spacelike boundaries in a spacetime where the outward null normals have vanishing expansion and the inward null normals have negative expansion - see ~\cite{Ashtekar2004} for a review. The geometry of these {\it apparent horizons}, which are distinct from the globally defined event horizons ~\cite{Booth2005}, correlates with the double-chirp pattern. 
Current simulation template banks ~\cite{SXS2019, Jani2016, Ferguson2023, Healy2022}, however, ack the type of horizon geometry data used in \cite{CalderonBustillo2020}, making this correlation difficult to study. Although recreating simulations across the full parameter space would require prohibitive computational resources, the analysis of existing simulations and waveform models can inform targeted follow-up studies.\par

%% Methodology
To this end, in this paper we parameterize the double-chirp behavior through $\kappa$, an effective metric that quantifies the post-merger content of a \bbh{} signal in time-frequency space, to enable a systematic investigation of the effects of the mass ratio $q$, the effective aligned spin $\xi$, and the precessing spin $\chi_p$ on the waveform morphology. This analysis establishes a foundation for future NR studies by linking waveform characteristics with the horizon geometry.
%
%% Paper structure
Section~\ref{morphology_basic} examines the morphology of \bbh{} systems producing double-chirp patterns and describes their parameterization using the continuous wavelet transform. Section~\ref{Skymaps} analyzes the distribution of post-merger emission on the final black hole's celestial sky across intrinsic parameter space. Section~\ref{disc} discusses results and future directions.\par

%%%%%%%%%%%%%%%%%%%%%%%%%%%%%%%%%%%%%%% 
\section{Double-chirp morphology \& parameterization}
\label{morphology_basic}
%%%%%%%%%%%%%%%%%%%%%%%%%%%%%%%%%%%%%%% 

%% define strain
The gravitational wave strain $h(t)$ can be expressed as:
\begin{align}
    h(t) = F_{+}\left(\theta, \delta, \psi\right) h_{+}(t) + F_{\times}\left(\theta, \delta, \psi\right) h_{\times}(t),
\end{align}
where $\left(F_{+}, F_{\times}\right)$ are detector response functions and $\left(h_{+}, h_{\times}\right)$ are polarizations, decomposed as:
\begin{align}
    h_{+}(t) + i h_{\times}(t) = \sum_{l \geq 2} \: \sum_{m = -l}^{m = l} \prescript{-2}{}{Y_{l,m}} \left(\iota, \phi\right) h_{l,m}(t).
    \label{sumovermodes}
\end{align}
Here, $\prescript{-2}{}{Y_{l,m}}$ are the spin-2 weighted spherical harmonic functions and $h_{l,m}(t)$ are the corresponding gravitational wave modes. The waveform morphology thus depends on the angular position of the observer $\left(\iota, \phi\right)$, while the luminosity distance $d_L$ only affects the amplitude.\par 

%% reference frames
Two reference frames characterize the \gw{} emission. The \emph{source frame} is centered on the instantaneous center of mass of the binary system, with $\hat{z}$ aligned with the angular momentum $\hat{L}$ at the reference frequency $f_{ref}$, and $\hat{x}$ pointing from $m_2$ to $m_1$; this is the frame in which the spin components are calculated.
The \emph{wave frame} is centered identically, with $\hat{Z}$ toward the observer and $\hat{X}$-$\hat{Y}$ defining the binary's sky plane. The inclination $\iota$ is the angle between the source frame $\hat{z}$ and the wave frame $\hat{Z}$ (0: face-on, $\pi/2$: edge-on, $\pi$: face-off), while $\phi$, commonly referred to as the reference phase, is the azimuthal angle.\par

%% analysis methodology
Our analysis uses waveforms generated via \texttt{pycbc}~\cite{pycbc} using NR templates~\cite{Jani2016, Ferguson2023} or approximant models~\cite{SEOBNRv4PHM}, with intrinsic parameters $\left(q, S_{1\, x,y,z}, S_{2\, x,y,z},\right)$ and observer orientation $\left(\iota, \phi\right)$. The total mass,  luminosity distance, and  starting frequency $f_{low}$ are fixed (i.e., a fixed initial separation in the binary~\cite{PhysRevD.52.821}), and the coalescence time $t_c$ is determined by the peak amplitude of the total waveform.\par

\begin{figure}
\includegraphics[width=0.4\textwidth]{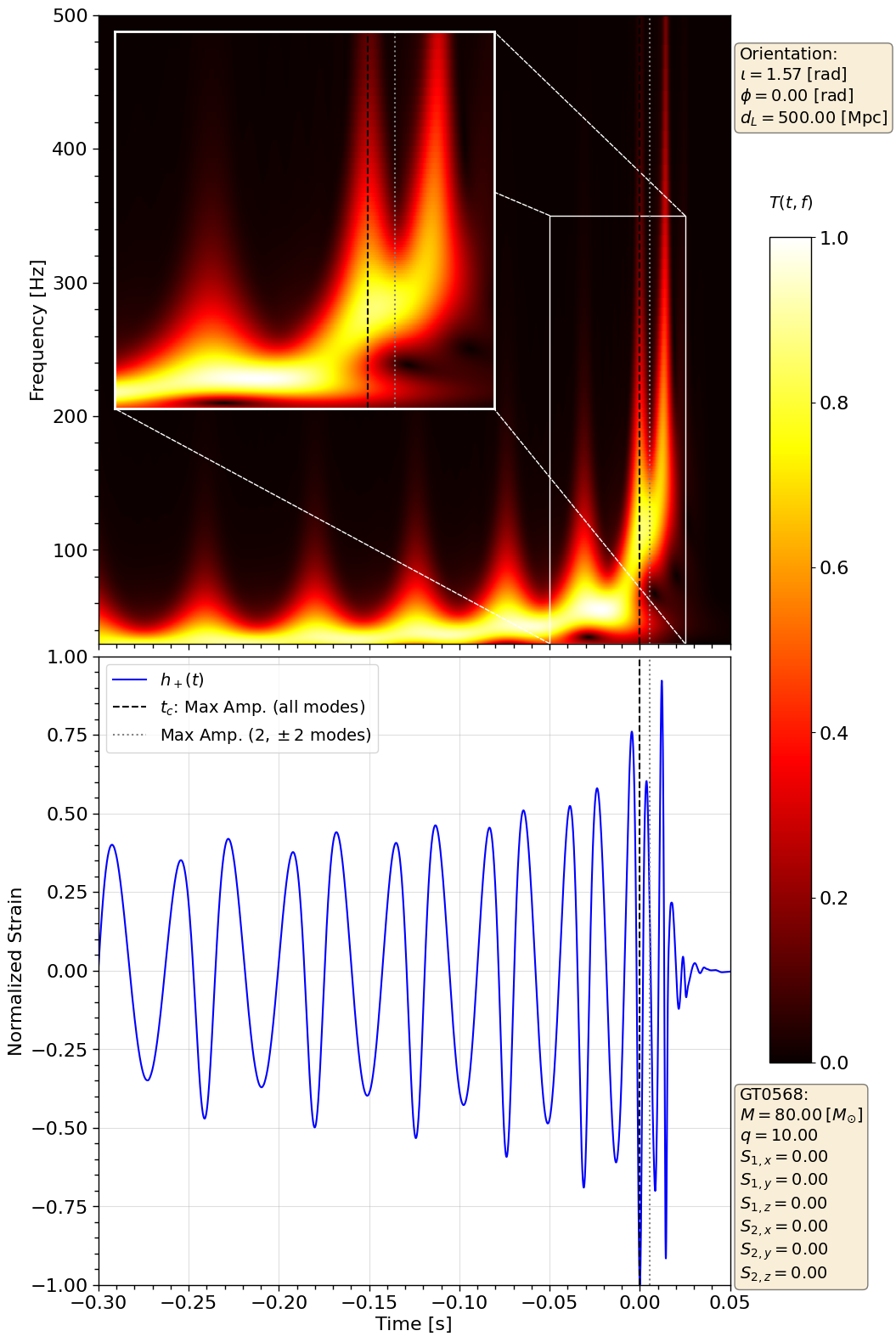}
\caption{Time-frequency structure in NR waveform GT0568, with $q = 10.0$ and $\abs{\mathbf{S}_1} = \abs{\mathbf{S}_2} = 0$, scaled to $M = 80 \: M_{\odot}$ at $d_L = 500$ MPc. The waveform is generated at a starting frequency of $f_{low} = 30\,$Hz, at the orientation $(\phi = 0.0, \iota = \pi / 2)$. The top panel displays the time-frequency map of the signal evolution. The bottom panel displays the real part of the (normalized) signal strain. The black dashed vertical line denotes the time of maximum amplitude for the total waveform, while the gray dashed vertical line denotes the time of maximum amplitude for the $\left(l,m\right) = \left(2,\pm2\right)$ modes. The inset highlights the distinct separation between the frequency spikes at the merger and in the post-merger.}
\label{GT0568_single}
\end{figure}

% introducing GT0568
For our initial investigation, we examine \nr{} waveform GT0568~\cite{Jani2016, Ferguson2023}, a zero-spin quasicircular \bbh{} merger with mass ratio $q = 10$ that includes spherical harmonic modes up to $l=8$; its waveform and time-frequency map are shown in Figure~\ref{GT0568_single}. We generate the free-space gravitational waveform (i.e., not projected onto a detector) using the \texttt{pycbc} package at a starting frequency of $f_{low} = 30\,$Hz, and scale the system to a total mass of $M = 80 \: M_{\odot}$ at $d_L = 500\,$MPc. The binary orientation relative to the observer is set at $(\phi = 0.0, \iota = \pi / 2)$; i.e. the observer is edge-on to the orbital plane. The time-frequency map is produced using the continuous wavelet transform (CWT) with a Morlet-Gabor wavelet basis at quality factor $Q=5.0$ with $n_f = 400$ logarithmically spaced frequencies from $f_{low}$ to $f_{max} = 500\,$Hz~\cite{Henshaw2024}.\par

The top panel of Fig.(\ref{GT0568_single}) shows, for this edge-on orientation, a double-chirp pattern with chirp-like features (henceforth referred to as spectral peaks) increasing through the inspiral phase. In the bottom panel, we see modulation in the inspiral waveform due to the mass asymmetry, and a distinctive amplitude peak in the post-merger that nearly reaches the same value as the overall peak amplitude.\par 

The waveform morphology and the appearance of secondary chirps depends on the observer orientation through the $\prescript{-2}{}{Y_{l,m}} \left(\iota, \phi\right)$ functions, as also demonstrated in~\cite{CalderonBustillo2020}. At face-on orientation $\left(\iota = 0\right)$ the time-frequency map manifests a single monotonic chirp, while near edge-on orientation $\left(\iota = \pi/2\right)$ additional chirping behavior appears in both the inspiral and postmerger. This effect is symmetric about the edge of the orbital plane, and as discussed in Sec.(\ref{skymap_section}) below this symmetry is crucial towards connecting the double-chirp pattern to the underlying system physics. At edge-on inclination, the dependence of the waveform morphology on $\phi$ is not symmetric and, in general, depends strongly on the system's mass ratio and spin as shown in Secs.(\ref{q_vary}, \ref{aligned_spin}, \ref{prec_spin}) below. However, in all cases there is a value of $\phi$ where the double-chirp pattern is maximized.\par

Understanding how this pattern depends on the intrinsic parameters of the system is the first step toward understanding how it connects to the geometry of the \fbh{} horizon. As such, we will begin by parameterizing the double-chirp pattern into metrics that can be compared across variations of both intrinsic parameters and observer orientations. We will then subsequently explore the conditions under which the double-chirp pattern appears most prominently, using waveforms from the MAYA \nr{}  catalog ~\cite{Jani2016, Ferguson2023} which include spherical harmonic modes up to $l=8$, and the SEOBNRv4PHM waveform model~\cite{SEOBNRv4PHM} that produces modes $\left(l, m\right) = \left(2,\pm2\right), \left(2,\pm1\right) , \left(3,\pm3\right), \left(4,\pm4\right), \left(5,\pm5\right)$.\par

\subsection{Parameterization}
\label{parameterization}

To quantify the correlation between the appearance of the double-chirp pattern, the observer's orientation, and the intrinsic system parameters, we have developed a comparative metric from the waveform's time-frequency representation. This metric is the sum of normalized continuous CWT coefficients $T(t,f)$ in a well-defined post-merger region in time-frequency space, which effectively is a quantitative measure of post-merger power. 

\begin{figure} [tb]
    \centering
    \includegraphics[width=0.45\textwidth]{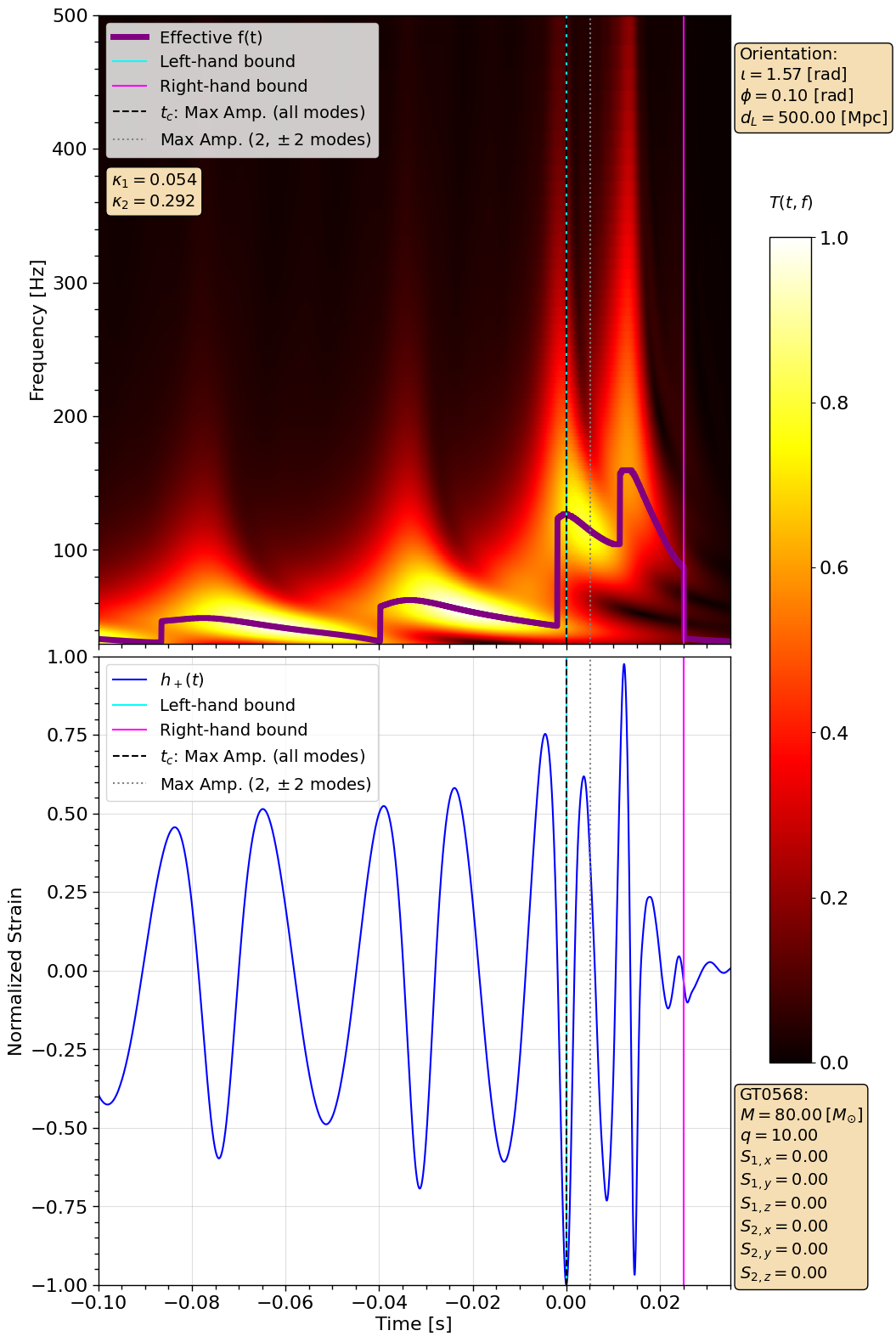}
    \caption{Spectral peaks in NR waveform GT0568 ($q = 10.0$, $\abs{\mathbf{S}_1} = \abs{\mathbf{S}_2} = 0$), scaled to $M = 80 \: M_{\odot}$ at $d_L = 500$ MPc, and generated starting from $f_{low} = 30\,$Hz. The top panel displays the time-frequency map of the signal evolution, created using CWT with a chirplet basis at d=0.2 \cite{Henshaw2024}. The bottom panel displays the real part of the signal strain. The purple line denotes the brightest pixel at each time step, tracing the effective one-dimensional $f(t)$. The cyan vertical line marks the \lhb{} of the $\kappa$ parameterization, and is coincident with the time of largest amplitude for the total waveform (all modes included), shown as a vertical black dashed line. The magenta vertical line marks the \rhb{} of the $\kappa$ parameterization, coincident with the sharp drop in the effective $f(t)$. The gray vertical dotted line denotes the time of largest amplitude for the $\left(l,m\right) = \left(2,\pm2\right)$ modes.}
    \label{param_GT0568}
\end{figure}

The analysis boundaries are established from characteristic features in the representations of the signal in both the time domain $h(t)$ and the time-frequency domain $T(t,f)$. A frequency-time trace $f(t)$ is constructed from maximum-intensity pixels in the time-frequency map at each discrete time interval, shown as a purple line in Fig.(\ref{param_GT0568}). One can see that this $f(t)$ trace has sharp separations between individual spectral peaks, both in the inspiral and in the post-merger, a feature that emerges from the use of chirplets in the CWT. For a discussion of how these features appear in the CWT map, we refer the reader to our companion work~\cite{Henshaw2024}. We bound the post-merger region between the time of maximum waveform amplitude (\lhb{}) and the sharp frequency drop-off (\rhb{}) in the $f(t)$ trace. For consistency across orientations, we fix the boundary separation $\kappa_w = RHB-LHB$ to the typical value of  $\kappa_w = 500$ samples ($\sim 0.03\,$s at a sampling rate of $16384\,$Hz), derived from a preliminary analysis of GT0568. In general $\kappa_w$ need only be large enough to encapsulate the post-merger frequency structure.\par

With this setup we define two metrics for post-merger content by summing over the $T(t,f)$ map:
\begin{align}
\kappa_1 = \frac{\sum_{i = LHB}^{RHB} \sum_{j=f_{low}}^{f_{max}} T(i, j)}{\sum_{i = 0}^{RHB} \sum_{j=f_{low}}^{f_{max}} T(i, j)},
\end{align}
is the fraction of post-merger to total signal power, and
\begin{align}
\kappa_2 = \frac{\sum_{i = LHB}^{RHB} \sum_{j=f_{low}}^{f_{max}} T(i, j)}{N_{pm}},
\end{align}
is the absolute post-merger power normalized by the number of pixels $N_{pm} = \kappa_w n_f$ in the post-merger region. For example, for the GT0568 waveform in Fig.\ref{param_GT0568}, $\kappa_1 \approx 0.052$ indicates that 5\% of the signal power is in the post-merger spectral peak, while $\kappa_2 \approx 0.289$ is the average post-merger power per pixel. Although the absolute value of $\kappa_2$ is less informative than $\kappa_1$, it serves as a comparative metric across variations in the intrinsic and extrinsic parameters.\par

The parameters $\kappa_1$ and $\kappa_2$ depend both on the physical parameters and on the analysis settings. The waveform generation parameters ($f_{low}$, sampling rate), the CWT parameters (Q factor), the reference frequency $f_{ref}$ and any other tunable settings must remain constant for meaningful comparisons. For example, increasing $f_{low}$ shortens the signal, yielding larger $\kappa$ values as the post-merger region comprises a greater proportion of the total signal.\par

To identify the system orientation that yields the strongest double-chirp pattern, we evaluate $\kappa_1$ across a uniform grid of $(\iota, \phi)$. In this work, we use a $16384\,$Hz sampling rate, impose a low-frequency cutoff  $f_{low}$ that varies based on available waveform length, and simulate sources at a luminosity distance of $500\,$Mpc. The CWT analysis uses a Morlet-Gabor wavelet basis and $n_f = 400$ logarithmically spaced frequencies from $f_{low}$ to $f_{max} = 500\,$Hz Since it is possible for a single wide feature in the $T(t,f)$ map to dominate the calculation of $\kappa_1$, it is important to use a sufficiently low $Q$ to resolve individual spectral features, so we use $Q=5.0$ for all CWT maps. The post-merger region width is fixed at $\kappa_w =500$ samples ($\sim 0.03\,$s). A grid is then constructed in the orientation parameters, taking uniform distributions on $\left(\iota, \phi\right)$. For each point on the grid, the waveform and its time frequency map are generated and $\kappa_1$ is evaluated. Note that because each grid point constitutes an independent binary orientation, and thus an independent observer location, each evaluation may be conducted in parallel. This scan is used to identify the orientation of the most prominent double-chirp pattern, corresponding to the highest value of $\kappa_1$ - note also that the orientation of largest $\kappa_1$ is also the orientation of largest $\kappa_2$. It is at this optimal orientation, henceforth denoted as $\left(\iota_\kappa, \phi_\kappa\right)$, that we may examine the correlation of the pattern with intrinsic parameters.\par

\subsection{Variation of mass ratio}
\label{q_vary}

The mass asymmetry in a \bbh{} system is known to introduce amplitude modulations during the inspiral at edge-on inclination, with the larger black hole dominating the signal (see e.g. ~\cite{Mehta2017}). Previous work has shown that the \bbh{} mass asymmetry also impacts its post-merger signal, with the prominence of the secondary spectral peak increasing with q~\cite{CalderonBustillo2020}.\par 

\begin{figure*}
    \centering
    \includegraphics[width=0.9\textwidth]{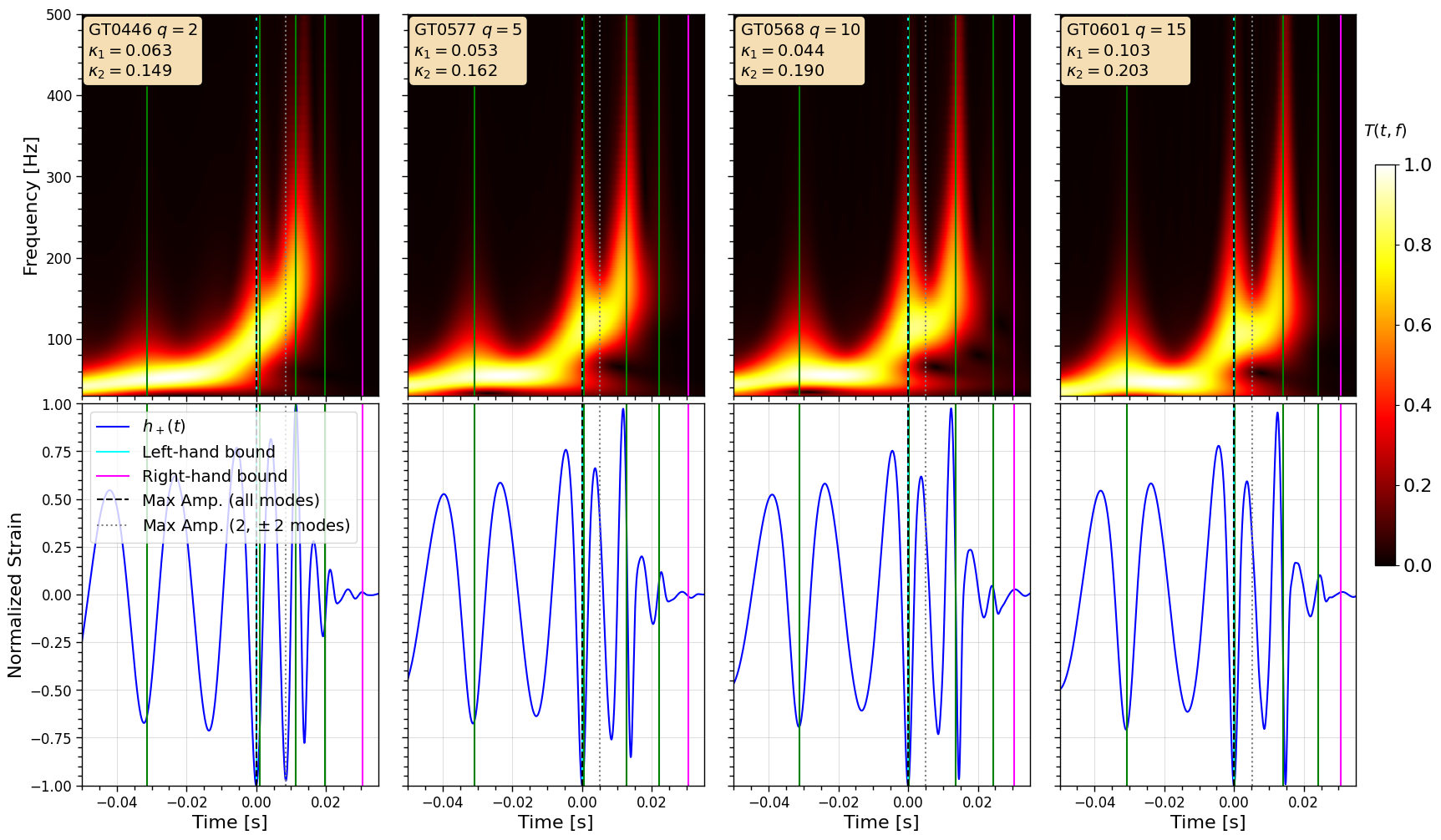}
    \caption{The relative prominence of spectral peaks as the mass ratio varies across waveforms GT0466, GT0577, GT0568, \& GT0601, spanning $q= \left[2, 5, 10, 15\right]$. Each waveform is scaled to $M = 80 \: M_{\odot}$ at $d_L = 500$ MPc, at optimal orientation $(\iota_\kappa, \phi_\kappa)$. Each waveform is generated at a starting frequency of $f_{low} = 30\,$Hz with the exception of GT0601, which starts at $f_{low} = 40\,$Hz. The top panels display the time-frequency map of the signal evolution. The bottom panels display the real part of the (normalized) signal strain. The cyan vertical line marks the \lhb{} of the $\kappa$ parameterization, and is coincident with the time of largest amplitude for the total waveform (all modes included), shown as a vertical black dashed line. The magenta vertical line marks the \rhb{} of the $\kappa$ parameterization. The gray vertical dotted line denotes the time of largest amplitude for the $\left(l,m\right) = \left(2,\pm2\right)$ modes. The green vertical lines denote the time location of spectral peaks.}
    \label{NR_qvary}
\end{figure*}

Fig.(\ref{NR_qvary}) illustrates this relationship through zero-spin NR waveforms GT0466, GT0577, GT0568, and GT0601, which span mass ratios q = [2, 5, 10, 15],  at the optimal orientation $\left(\iota_\kappa, \phi_\kappa\right)$. Fig.(\ref{NR_qvary}) shows that the mass ratio affects both the chirp pattern and the amplitude modulation, in agreement with the results of~\cite{CalderonBustillo2020}. Although the maximum values of $\kappa_1$ do not increase consistently with $q$, $\kappa_2$ increases with $q$, suggesting an increase in the post-merger emission. Higher mass asymmetry also enhances the spectral peaks in the inspiral.\par

Spectral peaks (marked by green vertical lines in Fig.(\ref{NR_qvary}) and discussed in App.(\ref{apndx_peak}) occur approximately every two waveform cycles during the inspiral, reflecting $f_{GW} \approx 2 f_{orb}$. This pattern changes in the post-merger, with spectral peaks occurring closer to 1.5 cycles after merger. Further details on this timing are contained in Sec.(\ref{morph_sum}) and physical implications are discussed in Secs.(\ref{Psi4_morph}, \ref{evidence}).\par

\subsection{Spin alignment}
\label{aligned_spin}

To analyze spin effects, we use the SEOBNRv4PHM approximant model~\cite{SEOBNRv4PHM}, validated for $q \leq 4$ and dimensionless spin magnitudes $\leq 0.9$, producing modes $\left(l, m\right) = \left(2,\pm2\right), \left(2,\pm1\right) , \left(3,\pm3\right), \left(4,\pm4\right), \left(5,\pm5\right)$.\par

\begin{figure*}
    \centering
    \includegraphics[width=0.9\textwidth]{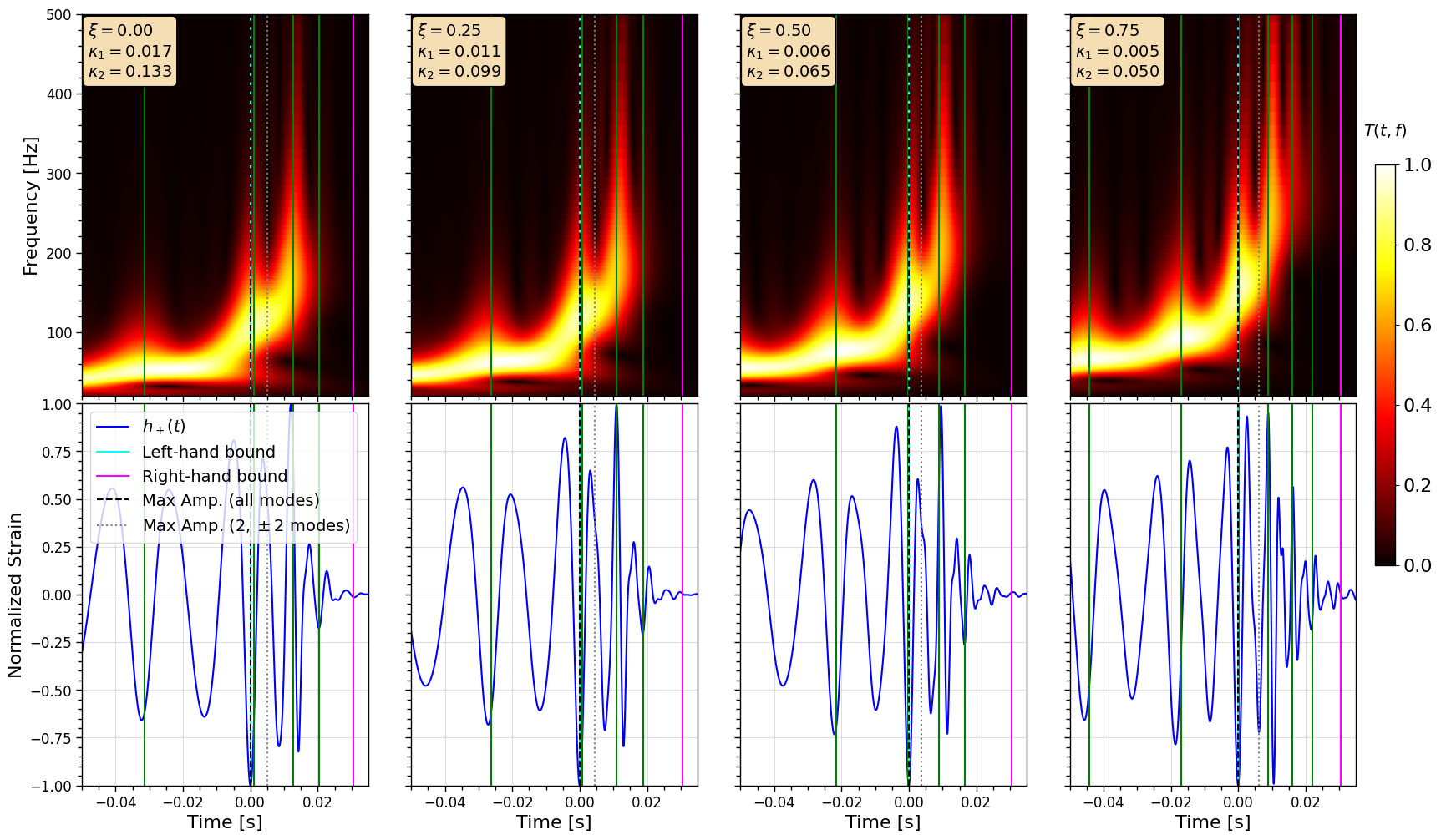}
    \caption{Variation of spectral peaks with effective aligned spin, using the SEOBNRv4PHM waveform model and spanning $\xi= \left[0.0, 0.25, 0.50, 0.75\right]$. Each waveform is scaled to $M = 80 \: M_{\odot}$ at $d_L = 500$ MPc, at optimal orientation $(\iota_\kappa, \phi_\kappa)$, and generated at a starting frequency of $f_{low} = 20\,$Hz. The top panels display the time-frequency map of the signal evolution. The bottom panels display the real part of the (normalized) signal strain. The cyan vertical line marks the \lhb{} of the $\kappa$ parameterization, and is coincident with the time of largest amplitude for the total waveform (all modes included), shown as a vertical black dashed line. The magenta vertical line marks the \rhb{} of the $\kappa$ parameterization, coincident with the sharp drop in the effective $f(t)$. The gray vertical dotted line denotes the time of largest amplitude for the $\left(l,m\right) = \left(2,\pm2\right)$ modes. The green vertical lines denote the time location of spectral peaks.}
    \label{SEOB_posxivary}
\end{figure*}

\begin{figure*}
    \centering
    \includegraphics[width=\textwidth]{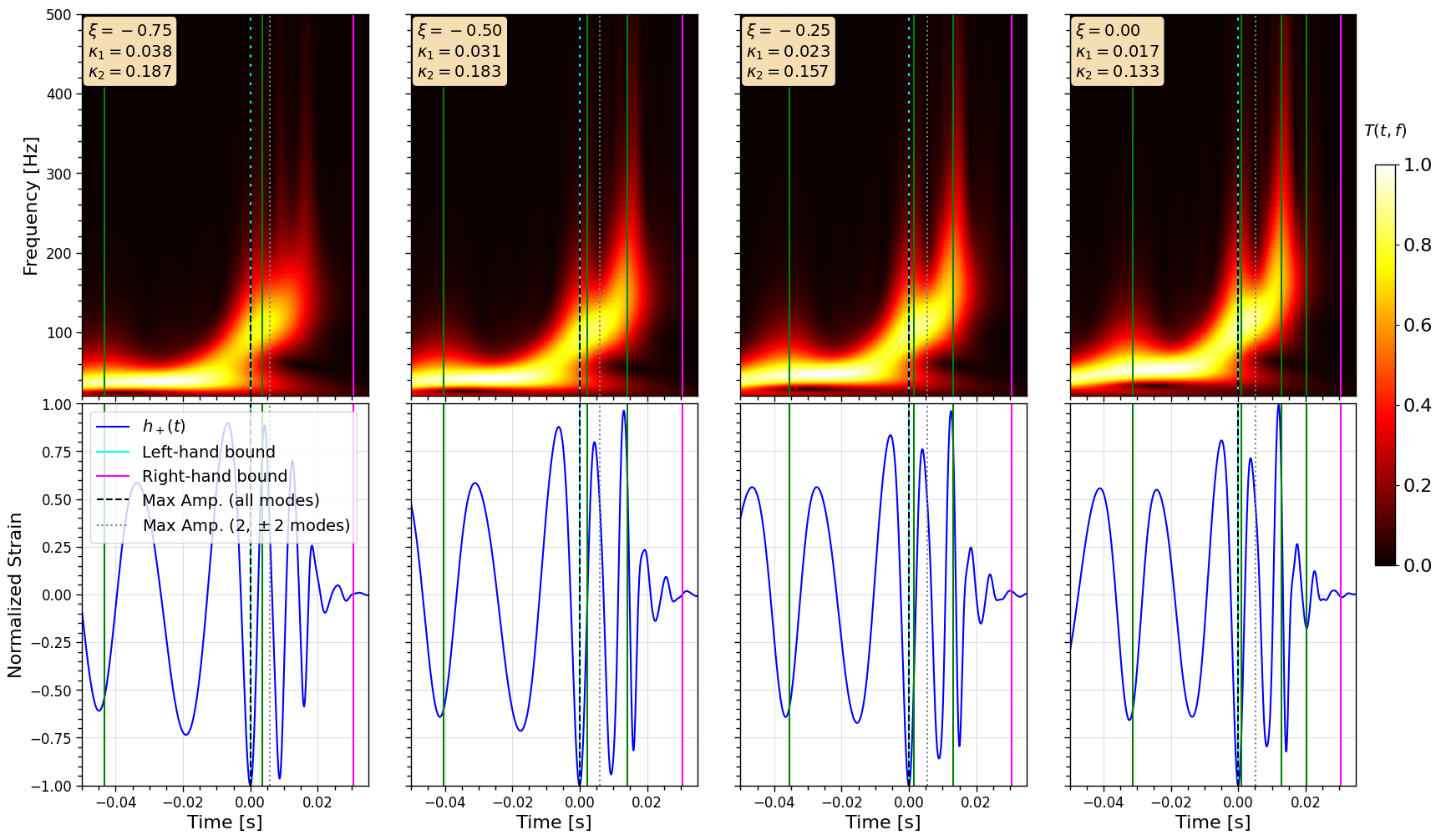}
    \caption{Variation of spectral peaks with effective aligned spin, using the SEOBNRv4PHM waveform model and spanning $\xi= \left[-0.75, -0.50, -0.25, 0.0\right]$. Each waveform is scaled to $M = 80 \: M_{\odot}$ at $d_L = 500$ MPc, at optimal orientation $(\iota_\kappa, \phi_\kappa)$, and generated at a starting frequency of $f_{low} = 20\,$Hz. The top panels display the time-frequency map of the signal evolution. The bottom panels display the real part of the (normalized) signal strain. The cyan vertical line marks the \lhb{} of the $\kappa$ parameterization, and is coincident with the time of largest amplitude for the total waveform (all modes included), shown as a vertical black dashed line. The magenta vertical line marks the \rhb{} of the $\kappa$ parameterization, coincident with the sharp drop in the effective $f(t)$. The gray vertical dotted line denotes the time of largest amplitude for the $\left(l,m\right) = \left(2,\pm2\right)$ modes. The green vertical lines denote the time location of spectral peaks.}
    \label{SEOB_negxivary}
\end{figure*}

For objects in a binary system with dimensionless spins $\vec{\chi}_i = c \vec{S}_i / (G m_i^2)$ and no transverse spin, the effective aligned spin parameter (sometimes referred to as $\chi_{eff}$) is:
\begin{align}
\xi \equiv \frac{1}{1+q}\left(q \chi_{1,z} + \chi_{2,z}\right),
\label{chi_eff}
\end{align}
where $\xi \in \left\{-1, 1\right\}$ indicates spin alignment (positive) or anti-alignment (negative) with $\hat{z}$. Aligned spins increase inspiral cycles, whereas anti-aligned spins reduce them~\cite{PhysRevD.52.821}. Figs.(\ref{SEOB_posxivary}, \ref{SEOB_negxivary}) show $\left(\iota_\kappa, \phi_\kappa\right)$ waveforms at $q = 4$, $M = 80\, M_{\odot}$, $d_L = 500\,$MPc for $\xi = \left[0.0, 0.25, 0.50, 0.75\right]$ and $\xi= \left[0.0, 0.25, 0.50, 0.75\right]$, respectively, achieved through appropriate selections of $\chi_{1,z}, \chi_{2,z}$.\par

Fig.(\ref{SEOB_posxivary}) demonstrates multiple effects of increasing the aligned spin $\xi$. Higher $\xi$ values add inspiral cycles, reducing the separation of spectral peaks (marked by green lines) and shifting the time-frequency map upward. Higher $\xi$ also corresponds to more prominent spectral peaks, particularly at merger, with a distinct additional post-merger spectral peak appearing at $\xi \ge 0.75$. Despite this additional post-merger structure, both $\kappa_1$ and $\kappa_2$ decrease with increasing $\xi$.\par

Fig.(\ref{SEOB_negxivary}) focuses on anti-aligned spins. As $\xi$ decreases, the separation between the final inspiral and the merger spectral peaks increases due to the reduced number of inspiral cycles, while the time-frequency map shifts to lower frequencies. The prominence of all spectral peaks decreases with decreasing $\xi$, becoming nearly indistinguishable at $\xi = -0.75$, where multiple weak spectral peaks emerge without clear physical correspondence. A distinct dual-peak structure only becomes apparent at $\xi = -0.50$, while $\kappa_1$ and $\kappa_2$ continue their inverse relationship with $\xi$. Further discussion on the correspondence of these patterns to the underlying physics may be found in Secs.(\ref{Psi4_morph}, \ref{evidence}).\par

\subsection{Precessing spins}
\label{prec_spin}

If either of the constituent spins is misaligned relative to the orbital angular momentum, the orbital plane will precess about the direction of total angular momentum. This precession behavior is known to accelerate throughout the inspiral, with timescale $t_{pre} \approx (r/M)^{5/2}$, where $r$ is the instantaneous separation between the two objects in the binary~\cite{Gerosa_2021}. The degree to which a system is precessing is often characterized by the parameter $\chi_p$~\cite{Schmidt_2015}:
\begin{align}
    \chi_p \equiv \max \left(\chi_1 \sin{\theta_1}, q\frac{4 q + 3}{4 + 3 q} \chi_2 \sin{\theta_2}\right),
\end{align}
where   $\theta_1$ and $\theta_2$ are the polar tilt angles of the spins with respect to the direction of orbital angular momentum, taken instantaneously at reference frequency $f_{ref}$.\par

To study the impact of precession  on the double-chirp pattern, we generate $q=4$ waveforms with $\chi_p = \left[0.0, 0.25, 0.50, 0.75\right]$, scaled to $M = 80 \: M_{\odot}$ at $d_L = 500$ MPc and $f_{ref} = 20\,$Hz. Note that given the high mass ratio and thus negligible contribution of the secondary spin, to scale the precession we increase only the value of $\chi_{1,y}$, with all other spin components held at zero. Fixing the spin orientation is crucial to isolate the impact of precession on the double-chirp pattern, as $\chi_p$ is degenerate across a selection of spin configurations and the exact spin orientation affects the nature of the remnant~\cite{Varma2019a}.\par 

\begin{figure*}
    \centering
    \includegraphics[width=\textwidth]{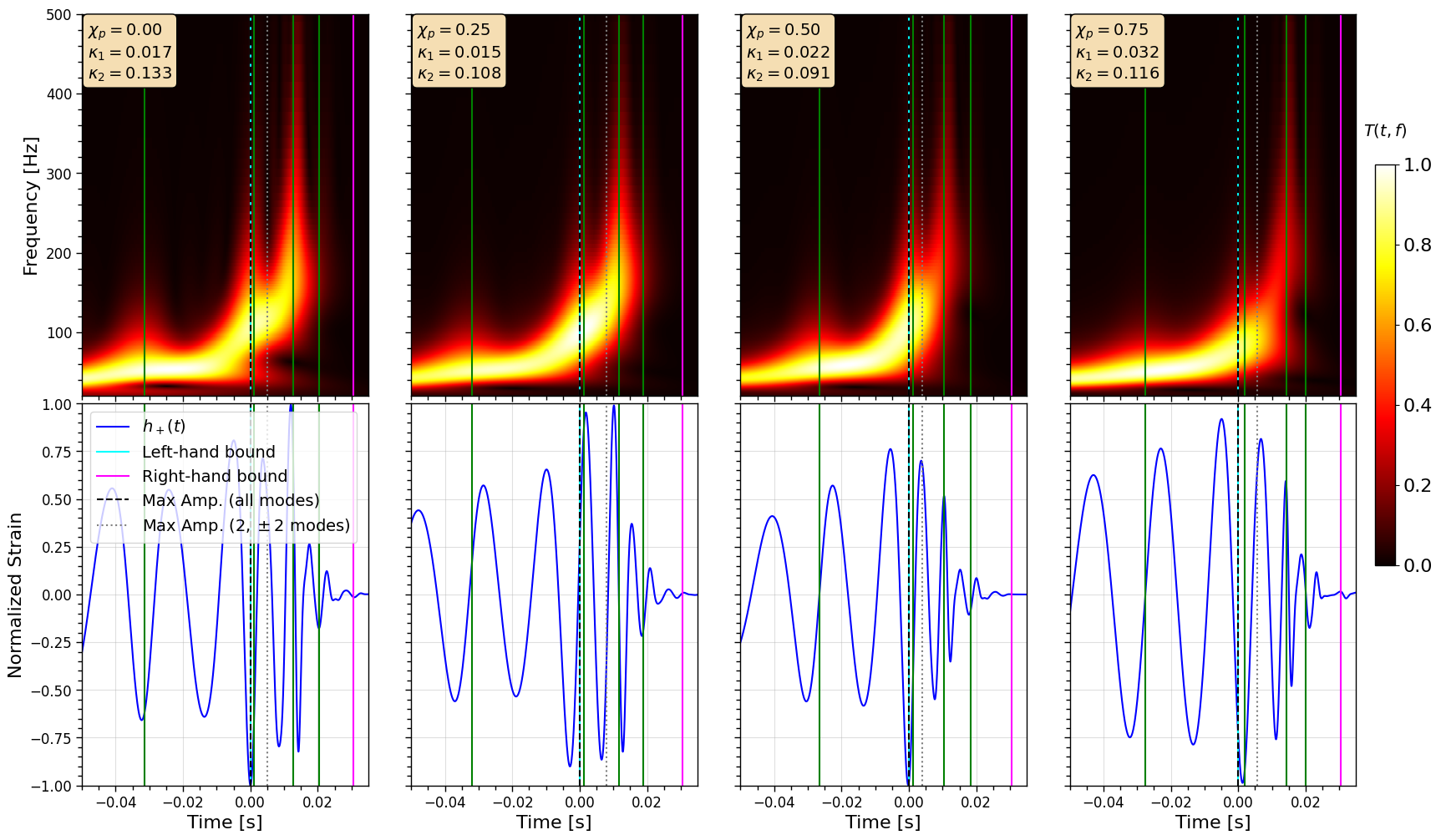}
    \caption{Variation of spectral peaks with precessing spin, using the SEOBNRv4PHM waveform model and spanning $\chi_p= \left[0.0, 0.25, 0.5, 0.75\right]$. Each waveform is scaled to $M = 80 \: M_{\odot}$ at $d_L = 500$ MPc, at optimal orientation $(\iota_\kappa, \phi_\kappa)$, and generated at a starting frequency of $f_{low} = 20\,$Hz. The top panels display the time-frequency map of the signal evolution. The bottom panels display the real part of the (normalized) signal strain. The cyan vertical line marks the \lhb{} of the $\kappa$ parameterization, and is coincident with the time of largest amplitude for the total waveform (all modes included), shown as a vertical black dashed line. The magenta vertical line marks the \rhb{} of the $\kappa$ parameterization, coincident with the sharp drop in the effective $f(t)$. The gray vertical dotted line denotes the time of largest amplitude for the $\left(l,m\right) = \left(2,\pm2\right)$ modes. The green vertical lines denote the time location of spectral peaks.}
    \label{SEOB_chipvary}
\end{figure*}

Fig.(\ref{SEOB_chipvary}) shows that precession complicates the waveform morphology. Spectral peaks in the inspiral are flattened, disappearing almost entirely for $\chi_p = 0.75$, similar to the anti-aligned spin case. However, the time separation between the pre-merger and the merger spectral peak decreases as the precession increases, which is more similar to the aligned spin case. We also see that the relative prominence of the merger versus post-merger spectral peaks does not follow a clear trend. The merger spectral peak is diminished at $\chi_p = 0.25$ and is barely distinguishable at $\chi_p = 0.75$, but is more prominent in the $\chi_p = 0, 0.5$ cases. The value $\kappa_1$ is similarly variable, but we see a downward trend in $\kappa_2$ as the precession increases. A discussion on the correspondence of these patterns to the underlying physics may be found in Secs.(\ref{skymap_section}, \ref{evidence}).\par

\subsection{Modal contribution}
\label{modes}

A common assumption when analyzing GW signals is to restrict waveforms to the dominant $(l,m) = (2, \pm 2)$ mode, but so far the waveforms used in this work have been composed of all the modes that their models allow. In the case of the NR waveforms, this is for all modes up to $l = 8$~\cite{Jani2016, Ferguson2023}. For SEOBNRv4PHM, this is the modes $(l,m) = (2, \pm 2), (2, \pm 1), (3, \pm 3), (4, \pm 4), (5, \pm 5)$~\cite{SEOBNRv4PHM}.\par

\begin{figure*}
    \centering
    \includegraphics[width=\textwidth]{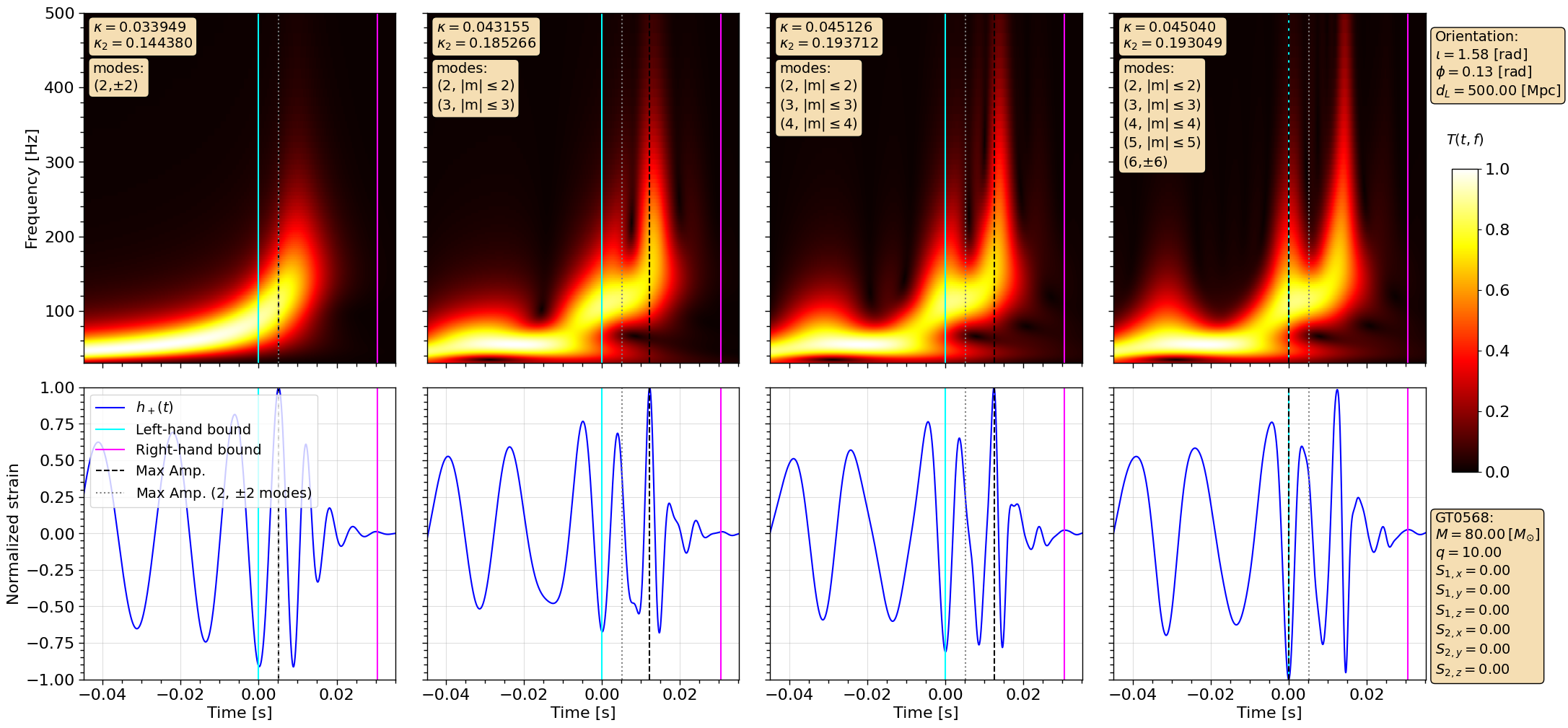}
    \caption{Double-chirp pattern appearance due to variation of modal composition for waveform GT0568 $\left(q = 10.0,\: \abs{\mathbf{S}_1} = \abs{\mathbf{S}_2} = 0\right)$, scaled to $M = 80 \: M_{\odot}$ at $d_L = 500$ MPc, and generated starting from $f_{low} = 30\,$Hz. The bottom panels display the real part of the (normalized) signal strain, and the top panels display the corresponding time-frequency map. From left to right, the columns display waveforms comprised of the $\left(l,m\right) = \left(2, \pm 2\right), \left(l\leq 3, \pm m\right), \left(l\leq 4, \pm m\right), \left(l \leq 5, \pm m\right) + (6, \pm 6)$ modes, where $m \in \left\{0,l\right\}$. The cyan dashed vertical lines denote the time of largest waveform amplitude for that column's mode case, and the magenta vertical lines denote the end of the $\kappa$ calculation window. The black dashed lines denote the time of largest waveform amplitude for the total mode waveform, and the grey dotted lines denote the time of largest waveform amplitude for the $\left(l,m\right) = \left(2, \pm 2\right)$ waveform.}
    \label{GT0568_modes}
\end{figure*}

To understand the contribution of specific \HM{s} to the double-chirp pattern, we will now consider waveforms with restricted modal composition. In Fig.(\ref{GT0568_modes}) we analyze the variation of modes in orientation $(\iota_\kappa, \phi_\kappa)$, starting from the $(l,m) = (2, \pm 2)$ mode, then adding $(3, |m| \leq 3)$, $(4, |m| \leq 4)$, and finally $(5, |m| \leq 5)$ and specifically $(6, \pm 6)$. In this figure, the time origin for each panel is set as the time of peak waveform amplitude for the total-mode waveform, highlighted by the vertical cyan line (that is, the \lhb{} in the $\kappa$ calculation). The vertical red dashed line shows the time location of the peak waveform amplitude for the restricted-mode waveform per panel, and the vertical gray dotted line shows the time location of the peak waveform amplitude for the $(l,m) = (2, \pm 2)$ mode.\par

Although the secondary spectral peak does not appear at all for the dominant $(2, \pm 2)$ mode alone,  the time-frequency structure begins to warp with the addition of the $(2, \pm 1)$ mode, particularly at low frequency. In the second panel of Fig.(\ref{GT0568_modes}) we see that with the addition of the $l=3$ modes (in particular the $(3, \pm 2)$ mode) the double-chirp pattern begins to appear; note that the time of peak waveform amplitude for the restricted waveform is centered on the secondary spectral peak. As modes are added, the peak amplitude shifts to the right, \textit{away} from the \lhb{}. This trend continues up to the addition of $(l, m) = (6, \pm 6)$, at which point the dominant amplitude jumps from the waveform peak that coincides with the secondary spectral peak to a trough coincident with the primary spectral peak, occurring before the $(2, \pm 2)$ mode peak. This amplitude peak is consistent with the \lhb{} for the remaining modes up to $l = 8$.\par

The rapid change in peak amplitude location upon addition of the $(l, m) = (6, \pm 6)$ mode does not mean that analysis of the double-chirp pattern requires modal content up to this level. Waveforms with modes below this still exhibit strong double-chirp patterns and will exhibit the same transitory jump in peak amplitude time, but at a different phase orientation. For example, a waveform with modes up to $l \leq 4$ is valid for the same suite of analysis, but, as discussed in Sec.(\ref{parameterization}) the determination of $t_c$ is dependent on the orientation $\left(\iota, \kappa\right)$. In Fig.(\ref{GT0568_modes}) we are fixing $\left(\iota, \kappa\right)$ as the determined value from the total-mode waveform. If one phase-advances the $l \leq 4$ waveform shown in the third panel, there will be a similar rapid transition in the time of peak amplitude at $\phi_\kappa \sim 336\degree$. It is for this reason that the determination of a global $t_c$ is deeply entangled with the observer orientation, and thus why sampling across the $\left(\iota, \phi\right)$ space is crucial towards analyzing the post-merger stage. Note that the misalignment of peak amplitude with the determined $t_c$ across orientations is a feature not a bug, and can be exploited to understand the distribution of post-merger power on the celestial sky of the \fbh{} - see Sec.(\ref{Skymaps}) below.\par

The impact of higher modes can also be quantified with $\kappa_2$. We find that $\kappa_2$ increases as modes are added, but levels out after $(l, m) = (4, \pm 4)$, as seen in the third and fourth panels of Fig.\ref{GT0568_modes}. At $l > 4$ the bulk characteristic of the double-chirp pattern is locked in, and additional higher modes have a minimal effect on the value of $\kappa_2$. 
Therefore, we recommend that $l = 4$ is the minimum mode level required to analyze the double-chirp pattern. We note that this requirement is met by waveform models such as SEOBNRv4PHM.\par

\subsection{Morphology Summary}
\label{morph_sum}

The double-chirp pattern is most prominent when the binary is oriented edge-on to the observer ($\iota = \pi / 2$ for a non-precessing system), with a phase angle $\phi$ that depends on the system's intrinsic parameters. We track this variation using the parameters $\kappa_1, \kappa_2$ that describe the post-merger power in the waveform's time-frequency map, and identify the orientation $\left(\iota_\kappa, \phi_\kappa\right)$ with the most prominent double-chirp pattern by finding the largest $\kappa_1(\iota, \phi)$. We have then studied the variation of intrinsic parameters $\left(q, \xi, \chi_p\right)$ at $\left(\iota_\kappa, \phi_\kappa\right)$ in Figs.(\ref{NR_qvary}, \ref{SEOB_posxivary}, \ref{SEOB_negxivary}, \ref{SEOB_chipvary}).\par 

For each of these cases, we have additionally tracked the time location of spectral peaks to understand their correlation to the intrinsic parameters of the system. These results are shown in Table(\ref{table:cs}), where $\Delta t_c$ is the difference in time of largest waveform amplitude between just the $\left(l,m\right) = \left(2,\pm 2\right)$ modes and the waveform generated with all available modes. The values in the $\Delta t_{chirp}$ column refer to the time spacing between spectral peaks, scanning from inspiral into the post-merger. Here we report always the final two pre-merger intervals followed by the post-merger intervals. In all cases the waveforms were scaled to $M = 80 \: M_{\odot}$ at $d_L = 500\,$MPc. All waveforms generated with the SEOBNRv4PHM approximate start at $f_{low} = 20\,$Hz, and the GT waveforms start at $f_{low} = 30\,$Hz with the exception of GT0601, which starts at $f_{low} = 40\,$Hz.\par

\begin{table*}
\centering
\renewcommand{\arraystretch}{1.3}  % Increase row spacing
\begin{tabular}{|>{\centering\arraybackslash}p{3cm}|c|c|c|c|c|c|c|c|}
\hline
\textbf{\large Model} & \textbf{\large $q$} & \textbf{\large $\xi$} & \textbf{\large $\chi_p$} & \textbf{\large $\iota_{\kappa} [\degree]$ } & \textbf{\large $\phi_{\kappa} [\degree]$} & \textbf{\large $\kappa_2$} & \textbf{\large $\Delta t_c$ [ms]} & \textbf{\large $\Delta t_{chirp}$ [ms]} \\ \hline \hline
GT0446       & 2 & 0.0 & 0.0 & 88.58 & 279.43 & 0.149 & 8.484 & $\{47.97, 32.47, 10.19, 8.30\}$ \\ \hline
GT0577       & 5 & 0.0 & 0.0 & 88.58 & 45.72 & 0.162 & 4.944 & $\{44.86, 31.43, 12.15, 9.40\}$ \\ \hline
GT0568       & 10 & 0.0 & 0.0 & 88.58 & 0.57 & 0.190 & 5.127 & $\{42.91, 31.43, 13.37, 10.74\}$ \\ \hline
GT0601       & 15 & 0.0 & 0.0 & 88.58 & 182.83 & 0.203 & 5.127 & $\{41.50, 30.94, 14.04, 10.07\}$ \\ \hline
SEOBNRv4PHM  & 4 & -0.75 & 0.0 & 88.58 & 0.57 & 0.222 & 5.676 & $\{77.03, 65.43, 47.00\}$ \\ \hline
SEOBNRv4PHM  & 4 & -0.50 & 0.0 & 88.58 & 279.43 & 0.217 & 5.981 & $\{70.37, 59.45, 42.54, 11.90\}$ \\ \hline
SEOBNRv4PHM  & 4 & -0.25 & 0.0 & 88.58 & 119.98 & 0.188 & 5.432 & $\{63.42, 52.92, 37.17, 11.47\}$ \\ \hline
SEOBNRv4PHM  & 4 & 0.0 & 0.0 & 88.58 & 262.87 & 0.133 & 5.066 & $\{46.88, 32.41, 11.78, 7.51\}$ \\ \hline
SEOBNRv4PHM  & 4 & 0.25 & 0.0 & 88.58 & 308.60 & 0.099 & 4.456 & $\{39.86, 27.22, 10.13, 7.93\}$ \\ \hline
SEOBNRv4PHM  & 4 & 0.50 & 0.0 & 88.58 & 268.55 & 0.065 & 3.784 & $\{32.71, 21.24, 9.34, 7.51\}$ \\ \hline
SEOBNRv4PHM  & 4 & 0.75 & 0.0 & 88.58 & 182.83 & 0.050 & 6.042 & $\{27.22, 17.09, 8.73, 7.02, 6.04\}$ \\ \hline
SEOBNRv4PHM  & 4 & 0.0 & 0.25 & 42.51 & 289.46 & 0.132 & 7.935 & $\{48.22, 33.20, 10.38, 7.26\}$ \\ \hline
SEOBNRv4PHM  & 4 & 0.0 & 0.50 & 86.86 & 148.34 & 0.112 & 3.906 & $\{43.09, 27.71, 9.28, 8.67\}$ \\ \hline
SEOBNRv4PHM  & 4 & 0.0 & 0.75 & 86.86 & 153.78 & 0.139 & 5.676 & $\{42.60, 29.54, 12.39, 6.47\}$ \\ \hline
\end{tabular}
\caption{Characteristics of systems analyzed in Sec.(\ref{morphology_basic}). $\left(\iota_\kappa,\phi_\kappa\right)$ gives the orientation that maximizes $\kappa_1$, $\kappa_2$ serves as a comparative metric for post-merger power, $\Delta t_c$ is the difference in time of largest waveform amplitude between just the $\left(l,m\right) = \left(2,\pm 2\right)$ modes and the waveform generated with all available modes, and the $\Delta t_{chirp}$ values refer to the time spacing between spectral peaks, scanning from inspiral into the post-merger.}
\label{table:cs}
\end{table*}

Note that in all cases $\Delta t_c$ is a positive number; the $\left(l,m\right) = \left(2,\pm 2\right)$ amplitude peak occurs after that of the full mode waveform, but before the first post-merger spectral peak. This further highlights the importance of mode selection in determining the time of merger and consequently the start of the post-merger regime. We see that as the mass ratio $q$ increases, the post-merger power $\kappa_2$ also increases. We see a trend of decreasing $\kappa_2$ as $\xi$ increases, indicating less raw post-merger power. Furthermore, as $\xi$ increases, the separation between spectral peaks decreases. We note that in all cases the intervals between spectral peaks in the inspiral encapsulate two wavelengths, in accordance with the $f_{GW} \approx 2 f_{orb}$ approximation. In the post-merger regime this approximation loses validity; the post-merger spectral peak intervals instead encapsulate roughly 1.5 wavelengths. As $\xi$ varies we note that the number of spectral peaks that appear in the post-merger likewise varies - the physical nature of this phenomenon will be explored in Sec.(\ref{Skymaps}) below.\par 

Finally, as the precessing spin $\chi_p$ increases, we get mixed results. The prominence of spectral peaks in the inspiral is flattened, but the time separation between spectral peaks decreases. The value of $\kappa_1$ increases with $\chi_p$, but the value of $\kappa_2$ shows no clear trend. These results highlight the difficulties in understanding how the precessing spin impacts both the post-merger waveform and its spectral content. This difficulty is due in part to a shortcoming of the waveform-centric morphology approach; by studying only the waveform at $\left(\iota_\kappa, \phi_\kappa\right)$, we have not considered the impact of changing intrinsics on the total distribution of post-merger power for all orientations. In the next section, we will expand this analysis to all orientations simultaneously for a given set of intrinsic parameters, and in doing so map the distribution of post-merger power on the \fbh{} sky.\par

\section{Distribution of post-merger power on the \fbh{} sky}
\label{Skymaps}

%The waveform morphology analysis in Sec.(\ref{morphology_basic}) considered the time domain waveform and its time-frequency representation at $\left(\iota_\kappa, \phi_\kappa\right)$, corresponding to the observer orientation relative to the merger that experiences the strongest double-chirp pattern. We have set $d_L$ as a constant, reducing the four-dimensional hypervolume of GW emission to three-dimensions. Fixing the orientation to $\left(\iota_\kappa, \phi_\kappa\right)$ is therefore examining only a one-dimensional slice of the three-dimensional emission volume. We will now relax this restriction and examine the distribution of post-merger emission across the full $\left( 0 \leq \iota \leq \pi, 0 \leq \phi \leq \pi/2\right)$ orientation space.\par

The study in  Sec.(\ref{morphology_basic}) focused on the optimal orientation for the double-chirp feature $\left(\iota_\kappa, \phi_\kappa\right)$;  we will now examine the distribution of post-merger emission across the full $\left( 0 \leq \iota \leq \pi, 0 \leq \phi \leq \pi/2\right)$ orientation space, at constant luminosity distance $d_L$.\par 

We do this with the same parallelization described in Sec.(\ref{parameterization}), with one significant change. The initial scan over grid points uniformly distributed in $\left( 0 \leq \iota \leq \pi, 0 \leq \phi \leq \pi/2\right)$ treats $t_c$ as an unbound local parameter, determined per-orientation in situ as the time of the largest waveform amplitude to identify $\left(\iota_\kappa, \phi_\kappa\right)$. Recognizing that $t_c$ must be a global parameter, we will now fix $t_c$ as the time of the largest waveform amplitude at the orientation $\left(\iota_\kappa, \phi_\kappa\right)$. With this fixed frame of reference in time, a subsequent grid of orientations may then be evaluated, and the full distribution of power may be studied in aggregate. For this second step, we will also switch from using the strain $h(t)$ to the $\Psi_4(t)$ scalar as described below.\par

\subsection{Post-merger $|\Psi_4(t)|$ in the orbital plane}
\label{Psi4_morph}

The Newman-Penrose scalar $\Psi_4(t) = \dv[2]{h^*(t)}{t}$  describes the Weyl tensor component related to outgoing transverse gravitational radiation; its magnitude $\abs{\Psi_4}$  is related to the intensity of gravitational radiation at a given point in spacetime~\cite{Newman1962}. It is shown in Fig. 2 of \cite{CalderonBustillo2020} that the post-merger distribution of $\abs{\Psi_4}$ in the orbital plane of a $q=3$ zero-spin \bbh{} simulation is correlated to the time separation between spectral peaks in the post-merger $\Psi_4$ waveform. Here, we will expand on that analysis in a backward fashion: we will start from waveform models that produce $h(t)$ and analyze the corresponding distribution of $\Psi_4 (t)$ in the binary orbital plane.\par

\begin{figure*}
    \centering
    \includegraphics[width=\textwidth]{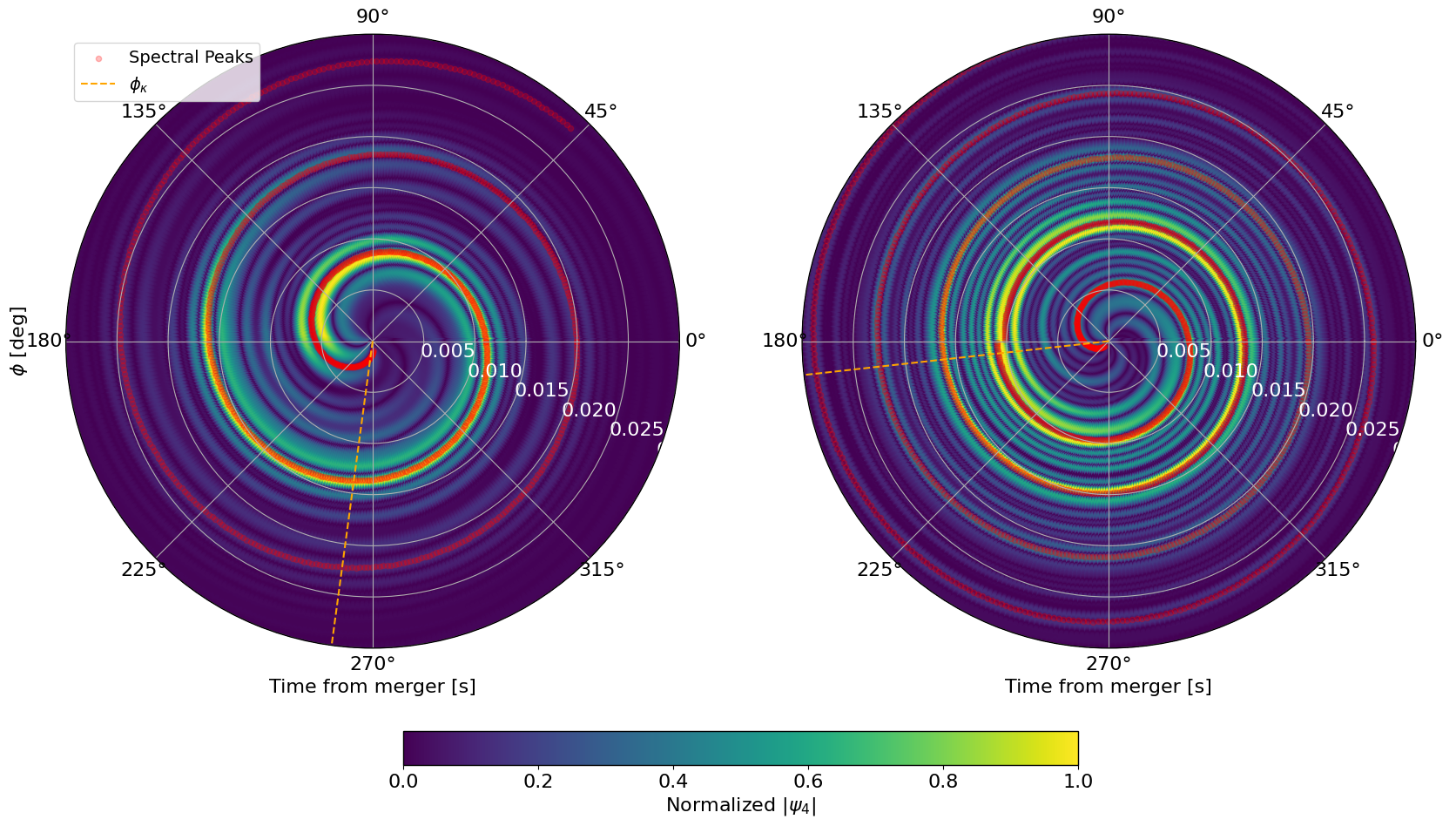}
    \caption{The distribution of spectral peaks in the orbital plane from the $\Psi_4(t)$ representation of SEOBNRv4PHM waveforms. The left-hand image shows a zero-spin system, while the right-hand image shows a system with $\xi = 0.75$. The radial axis shows the time following the merger, and the angular axis spans the phase angle $\phi$. The color bar shows the normalized absolute value of $\Psi_4$ per point. The red dots show the location of spectral peaks in the waveform. The orange dashed lines show the $\left(\iota_\kappa, \phi_\kappa\right)$ orientation. The sample size for each is N = 360 orientations.}
    \label{Psi4_aligned}
\end{figure*}

Fig.(\ref{Psi4_aligned}) shows the results of this computation for both the $\xi = 0$ (left) and $\xi = 0.75$ (right) systems modeled with SEOBNRv4PHM. In these diagrams, the radial axis represents the time following the merger, and the angular axis spans the phase angle $\phi$. The angular resolution is determined by the number of orientations sampled ($N=360$ values of $\phi$), while the temporal resolution is limited only by the sampling rate of the waveform ($16384\,$Hz). The color bar describes the value of $|\Psi_4|$ per point, normalized by the largest value. The red dots show the time location of the post-merger spectral peaks per orientation.\par

In the $\xi=0$ case, there is good agreement between the pattern traced by the spectral peaks and the underlying $\Psi_4$ distribution, although the peak value of $|\Psi_4|$ does not align with $\phi_\kappa$. These two quantities are obtained in fundamentally different ways: $\Psi_4$ is obtained directly from the waveform data, requiring only two derivatives of the strain $h(t)$, while spectral peaks are obtained by passing the $\Psi_4$ waveform through the CWT to create a time-frequency map, as well as a peak-finding algorithm - see App.(\ref{apndx_peak}). The general agreement between these two distributions confirms the peaks are physical, rather than numerical artifacts from the CWT.\par

Furthermore, this result supports the analysis shown in Fig. 2 of~\cite{CalderonBustillo2020}, where it is demonstrated through NR that there is a connection between the structure of GW emission and the time-frequency morphology of the signal. As in that analysis, we show here that both the number and intensity of spectral peaks experienced will vary depending on the observer orientation. Based on this diagram, an observer at $\phi = 90\degree$ will experience an initial very strong spectral peak followed by two progressively weaker ones. A different observer at $\phi = 0\degree$ will experience a weaker initial spectral peak, their secondary will be weaker still, and may not see a third as the signal strength is too weak so far from the merger time.\par

The right panel of Fig.(\ref{Psi4_aligned}) shows the same analysis for the $\xi = 0.75$ case. One can see that with the addition of aligned spin, there are many more cycles in the post-merger stage. In comparing these two images, consider, for example, an observer oriented at $\phi = 0$. In the same time span of $0.03\,$s, an observer of the $\xi = 0$ system would see two post-merger spectral peaks, while an observer of the $\xi = 0.75$ system would see four. Given the correlation between the spectral peaks and the passage across the observer's line of sight of sharp curvature on the \fbh{} horizon as proposed in \cite{CalderonBustillo2020}, we should expect that the aligned spin $\xi$ will affect the distribution of spectral peaks in the post-merger. Specifically, on the same time scale, a system with positive $\xi$ will have a greater density of spectral peaks, corresponding to more line of sight passages due to the reduction in orbital period. Further discussion on the physical nature of this effect may be seen in Sec.(\ref{disc}) below.

\subsection{Distribution of $\kappa$ across the \fbh{} celestial sky}
\label{skymap_section}

\begin{figure*}
    \centering
    \includegraphics[width=\textwidth]{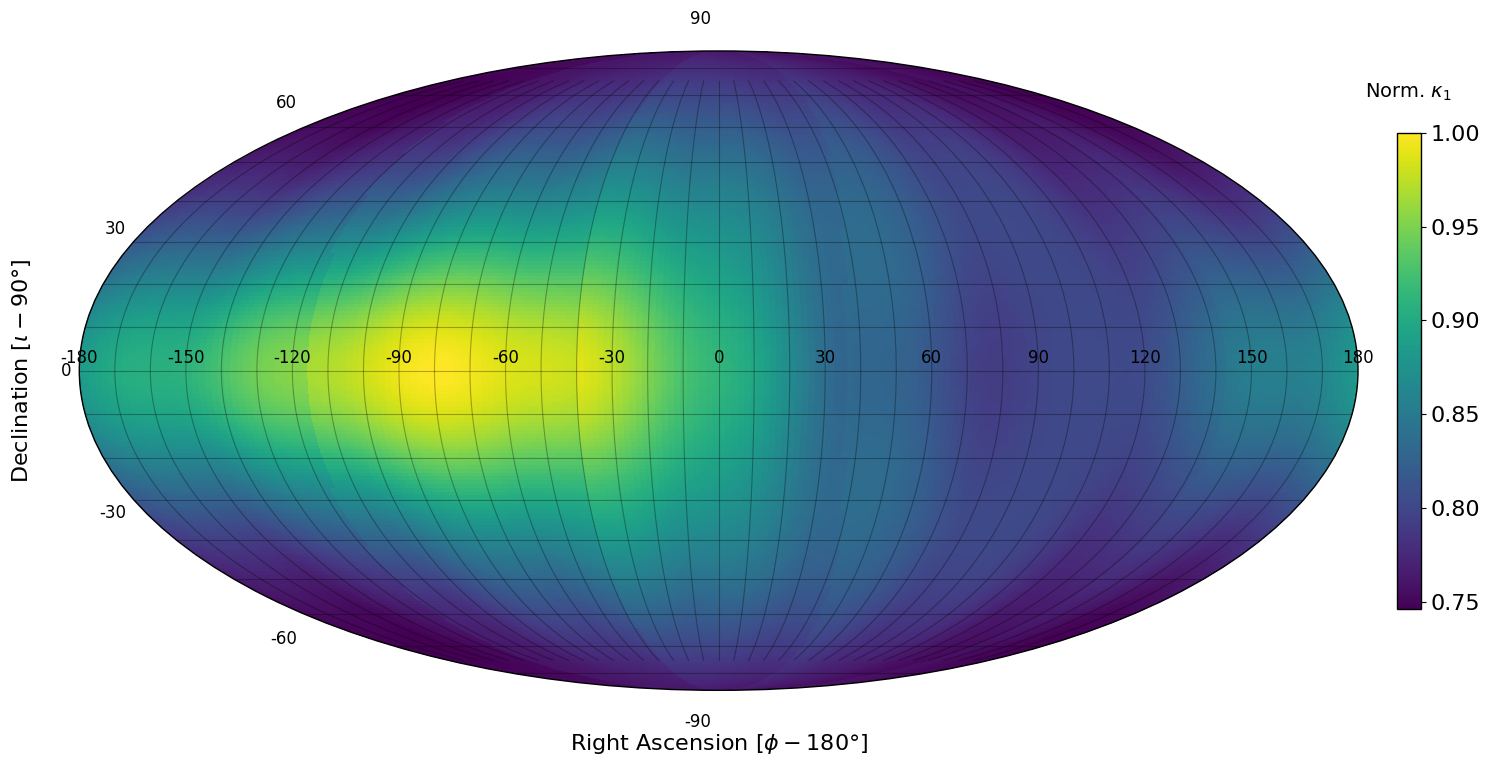}
    \caption{The distribution of post-merger radiative power as measured by $\kappa_1$ across the celestial sky of the simulated merger event GT0568, holding $t_c$ as a fixed global parameter. The value of $\kappa_1$ is evaluated at $N = 40,000$ grid points distributed uniformly in $\iota \in \left\{0, \pi\right\}$, $\phi \in \left\{0, 2\pi\right\}$, then normalized by the value of $\kappa_1$ at orientation $\left(\iota_\kappa, \phi_\kappa\right)$.}
    \label{GT0568_Psi4_skymap}
\end{figure*}

Next, we study the distribution of post-merger emission across the full $\left( 0 \leq \iota \leq \pi, 0 \leq \phi \leq \pi/2 \right)$ orientation space, which maps the celestial sky of the \fbh{} for observers at fixed $d_L = 500\,$ Mpc. Fig.(\ref{GT0568_Psi4_skymap}) is the realization of this method for GT0568. In this image, we plot the distribution of $\kappa_1$ as a Mollweide projection, with right ascension given by $\phi - 180\degree$ and declination given by $\iota - 90\degree$. The zero-latitude equator thus represents observer orientations edge-on to the orbital plane. The color bar displays the normalized values of $\kappa_1$.\par

In this visualization, the distribution of power on the \fbh{} sky is smooth and continuous, and it conforms to the expected behavior based on the morphology study in Sec.(\ref{morphology_basic}). The distribution is symmetric with respect to the zero-latitude line, corresponding to the edge-on orientation. Starting from a longitude of zero and tracing along the latitude lines, one can see that the value of $\kappa$ gradually increases, peaks at a latitude of $\sim -80\degree$ and then gradually decreases. The bulk of the post-merger power is concentrated in a tight area spanning $\sim 90\degree$ in longitude and $\sim 40\degree$ in latitude. Note also that there is a corresponding region of low-power on the opposite side of the sky, and even deeper voids on the poles.\par

The symmetry in Fig.(\ref{GT0568_Psi4_skymap}) about the zero-latitude line is crucial to understand how the distribution of post-merger power relates to the underlying physics of the system. GT0568 is a zero-spin simulation, so over the course of the inspiral the orbital plane does not change its orientation. For such systems, the double-chirp pattern is strongest at edge-on inclination $(\iota = \pi/2)$, and as such, we should expect the distribution of $\kappa$ for some constant $\phi$ (i.e. along a line of latitude) to be symmetric and peak at $\iota = \pi/2$. Furthermore, we should expect the introduction of aligned spin to maintain this symmetry, as the orientation of the orbital plane will remain fixed.\par

However, we may conjecture that this will not be the case for a system with precessing spins. If either of the spins $\left(\vec{\chi}_1, \vec{\chi}_2\right)$ is misaligned relative to the direction of the orbital angular momentum $\hat{L}$, then $\hat{L}$ will precess about the direction of total angular momentum $\hat{J}$. As such, the orientation of the orbital plane will vary over the course of the inspiral. Therefore, compared to its starting orientation at $f_{ref}$, the inclination value that describes the edge-on observer position will generally not be $\iota = \pi/2$ at the time of merger.\par

Additionally, if we follow the ideas presented in~\cite{CalderonBustillo2020}, the spectral peaks in the post-merger stage are correlated to a region of sharp curvature on the equator of the \fbh{} horizon that passes across the observer's line of sight. In this paradigm, we may conjecture that a precessing system should trace a more complicated pattern on the \fbh{} sky, as there must be correlation between the system dynamics, the deformation on the \fbh{} horizon, and the outgoing post-merger radiation. Following this conjecture, we should expect that as the precessing spin $\chi_p$ at $f_{ref}$ increases, the resultant distribution of postmerger power will break the symmetry about the zero-latitude line. \par

\begin{figure*}
    \centering
    \includegraphics[width=\textwidth]{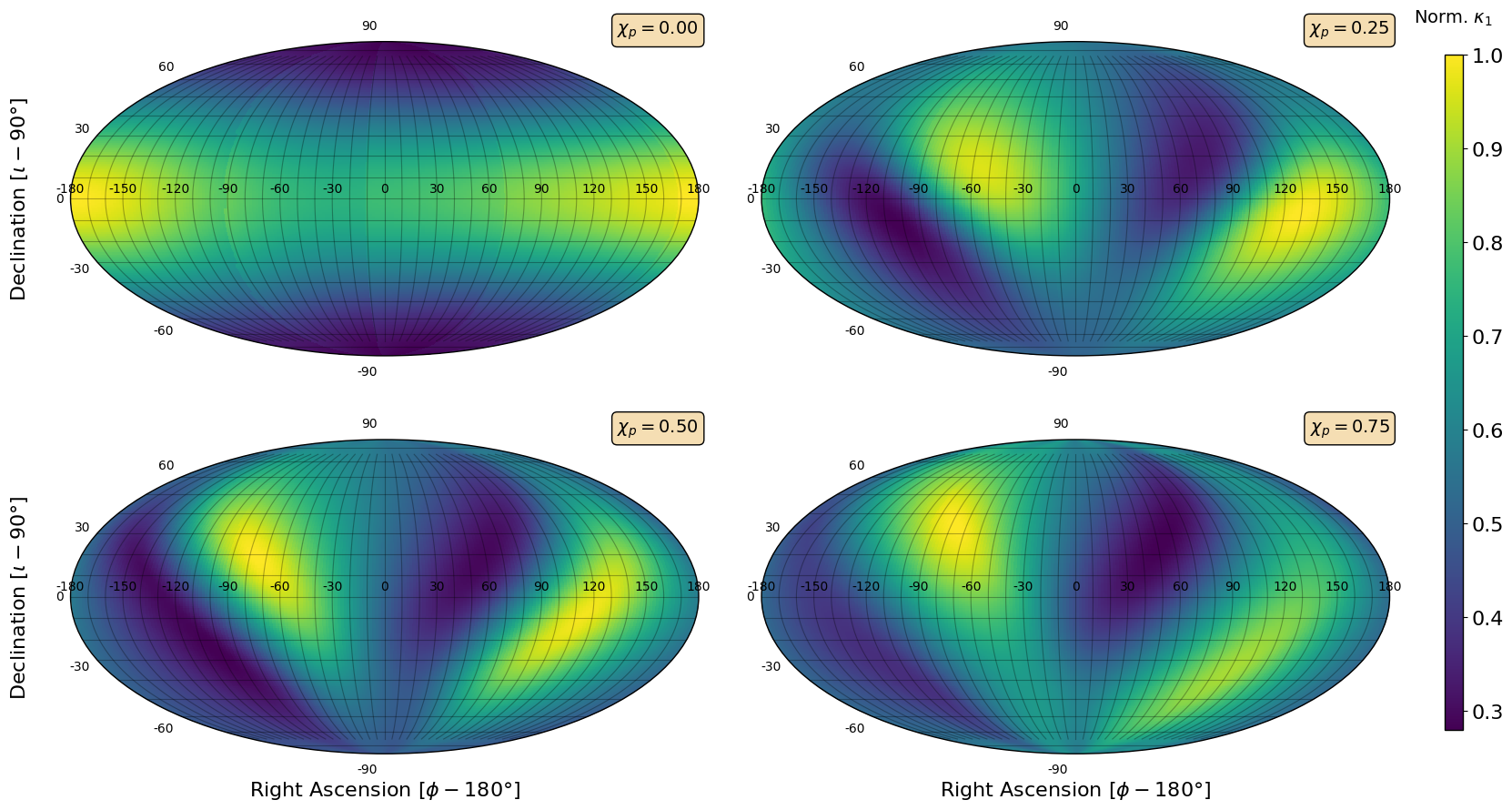}
    \caption{The distribution of post-merger power as measured by $\kappa_1$ across the celestial \fbh{} sky for different values of $\chi_p$ with all other intrinsics held fixed. In each case $t_c$ is also held as a fixed global parameter. The value of $\kappa_1$ is evaluated at $N = 40,000$ grid points per system distributed uniformly in $\iota \in \left\{0, \pi\right\}$, $\phi \in \left\{0, 2\pi\right\}$, then normalized by the value of $\kappa_1$ at orientation $\left(\iota_\kappa, \phi_\kappa\right)$.}
    \label{skymap_chipvary}
\end{figure*}

A demonstration of this effect is shown in Fig.(\ref{skymap_chipvary}) using waveforms generated with SEOBNRv4PHM, where one can see that increasing the system's precession has a profound impact on the distribution of post-merger power. In this image, the top-left panel shows the zero-spin case and bears many similar features to the GT0568 case. The distribution is symmetric about the zero-latitude line and consists of a concentrated bright swatch, a region of low power on the opposite end of the sky, and voids on the two poles.\par

Proceeding to the top-right panel with $\chi_p = 0.25$, we now see that this pattern has already changed dramatically. The symmetry across the zero-latitude line has been broken, and the bright swatch of outgoing power has split into two distinct regions. The two voids have also migrated; the right-hand void was the north pole, the left-hand the south. Although not apparent from this diagram, this change occurs very early in the continuum of progression over the range $\chi_p \in \left\{0,\right\}$, becoming distinctive at $\chi_p \approx 0.05$. By $\chi_p$ = 0.25 the sky pattern is mainly locked in, with additional spin magnitude serving to transfer more power from the right-hand to the left-hand swatch.\par 

% need to transition into discussion now

\section{Discussion}
\label{disc}

In this work, we have extended the investigations of \cite{CalderonBustillo2020} into the time-frequency structure of \gw{} signals from highly-inclined asymmetric BBH systems. We have demonstrated for the first time how the double-chirp pattern may be analyzed with a pure waveform approach. By parameterizing the post-merger power (see Sec.(\ref{parameterization}), we have developed a method to determine the observer orientation $\left(\iota_\kappa, \phi_\kappa\right)$ for which the double-chirp pattern is maximized given the intrinsic instantiation of an asymmetric mass ratio binary and a model that can simulate its gravitational waveform across orientations.\par

Using the SEOBNRv4PHM approximant model, we have explored the double-chirp pattern across a broad parameter space, including both aligned and misaligned spins. While the effective aligned spin $\xi$ is known to influence the binary's evolution and inspiral frequency~\cite{PhysRevD.52.821}, its impact on the post-merger waveform's time-frequency map was previously unexamined. Our results show that increasing $\xi$ reduces the separation between spectral peaks and can induce additional spectral peaks, reflecting a similar influence to that observed during the inspiral. We have further shown that the precessing spin $\chi_p$ has a significant effect on the distribution of post-merger power across the celestial sky of the \fbh{}, breaking the equatorial symmetry seen in non-precessing systems.\par

We next explore the physical meaning of these findings within the framework of \cite{CalderonBustillo2020}, which demonstrated a correlation between post-merger spectral peaks and the passage of sharp curvature on the \fbh{} apparent horizon across the observer's line of sight. In this way the geometry of the \fbh{} becomes a potential observable in GWs that reach us on Earth, suggesting that it may be possible to infer features related to the horizon geometry from analysis of GW data. To understand this possibility and all that it implies, we begin first with the physical origin of the post-merger spectral peaks.\par

\subsection{Evidence of physical origin}
\label{evidence}

Do the post-merger spectral peaks that we have analyzed in this work correspond to the evolution of features on the \fbh{} horizon? In this paper, we have investigated two primary pieces of evidence: (1) the morphology of spectral peaks at the orientation of strongest double-chirp, and (2) the distribution of post-merger power across the \fbh{} sky. To understand these results, we will first present a conjectural analogy regarding the continuation of BH trajectories from the inspiral into the post-merger.\par

Table \ref{table:cs} shows that $\Delta t_{chirp}$ continues to decrease from the inspiral into the post-merger. In the inspiral this interval agrees with the $f_{GW} \approx 2 f_{orb}$ approximation, suggesting that spectral peaks correspond to passage of the lighter BH $m_2$ across the observer's line of sight, with the following trough representing the likewise passage of the heaver BH $m_1$. To build intuition, we will now conjecture that this paradigm extends beyond the formation of the common horizon into the post-merger. One can imagine that at the instant of merger the point-like singularities containing $m_1, m_2$ simply proceed on their current trajectories and continue to inspiral within the newly formed common horizon. Note that this is explicitly not the case for the original horizons of $m_1, m_2$; their evolution is more complex~\cite{Pook-Kolb2021a, Pook-Kolb2021b, Evans2020, Booth2020}.\par 

Under this paradigm the singularity trajectories can be considered the primary generators of gravitational energy surrounding the entire inspiral, merger, post-merger transition. The spacetime curvature they induce on the near-merger region is the source of the horizon curvature, the outgoing radiation that reaches distant observers, and the ingoing radiation that enters the \fbh{} horizon. Note that there is an important subtlety here; information from within the \fbh{} horizon cannot be explicitly communicated to the exterior. One cannot attribute cause to the actions of objects that are causally disconnected from the exterior universe. However, as there is a common origin of the local spacetime curvature, \fbh{} horizon geometry, and outgoing radiation, correlation is possible. For the purposes of understanding if such correlation exists and it is observable, we will use this ``continued trajectory" paradigm.\par

Let us now examine the natural predictions for each piece of evidence that follow from the above conjecture, starting with the impact of aligned spin. The aligned spin~\cite{PhysRevD.52.821}, adds or subtracts cycles per second from the inspiral frequency for $+\xi$ or $-\xi$, respectively. We should therefore expect similar behavior to continue from the inspiral into the post-merger, with additional $+\xi$ adding cycles per second to the $m_1, m_2$ trajectories. If there is correlation, then within the same time frame we should expect a distant observer to see more spectral peaks in the post-merger of a $\xi>0$ binary than a $\xi =0$ binary.\par

We see this exactly in both the 2D post-merger distribution (Fig.(\ref{Psi4_aligned})) and the 1D waveform representation (Fig.(\ref{SEOB_posxivary})). In comparing the $\xi = 0.0$ and $\xi = 0.75$ systems, one can see that for a given observer orientation over the same post-merger time interval of $0.025\,$s the $\xi = 0.75$ system produces roughly twice the number of spectral peaks. Note that we should not expect the $f_{GW} \approx 2 f_{orb}$ approximation to hold into the post-merger. Furthermore, we do not necessarily need to establish the exact mapping of orbital evolution to gravitational wave frequency in this regime; we need only establish that such a mapping exists and that its specific realization depends on the intrinsic parameters of the system. We additionally see that there is good agreement between the time spacing of post-merger spectral peaks and the distribution of $|\Psi_4|$ within the orbital plane. However, one will note that particularly for the $\xi = 0.75$ system there are features in the underlying $|\Psi_4|$ distribution that are not perfectly tracked by the spectral peak trace. We speculate that this is due in part to the difficulty in tracking the spectral peak location - see App.(\ref{apndx_peak}) - and hope to improve on this in future work.\par

For systems with zero or purely aligned spin, the orientation of the orbital plane will remain fixed over the course of the coalescence. As seen in Fig.(\ref{GT0568_Psi4_skymap}), this leads to a symmetrical distribution in post-merger power across the celestial sky of the \fbh{}, straddling the equator that represents the edge-on $\left(\iota = \pi/2\right)$ observer orientation. If instead the system instead has misaligned spin, then the orientation of the orbital plane will precess about the direction of total angular momentum. In this case $m_1, m_2$ are not fixed to a plane and will inspiral through more complicated helical trajectories. In the continued trajectory paradigm, this means that we should expect the post-merger behavior to trace a more complicated pattern across the \fbh{} sky, and specifically to break the symmetry about the equator.\par

We see this happen in Fig.(\ref{skymap_chipvary}), where the addition of precessing spin through increase of $\chi_{1,y}$ causes the symmetry in the $\chi_p=0$ distribution to break as early as $\chi_p=0.25$. The resultant pattern has two hot spots and two cold spots. Notice that symmetry is not entirely lost; with induced precession there is now a longitudinal reflection symmetry between the hot spots and cold spots. This is not unexpected, as we should expect symmetry to be preserved under some coordinate transformation, as precessing systems can be modeled as a twisted-up version of non-precessing coordinates~\cite{Schmidt2012}. Further increase in spin magnitude shifts which hot/cold spot pair contains the majority of the power. Additionally, as the spin magnitude increases the line of symmetry moves to the left, indicating a dependence on the final phase orientation of the binary due to the increased rate of precession.\par 

These two pieces of evidence support the physical reality of the correlation proposed in \cite{CalderonBustillo2020}. However it is important to stress that these results do not attribute causation of post-merger spectral peaks to the horizon geometry. Even the attribution of correlation to the horizon geometry is tenuous, as with waveforms alone we cannot directly test such correlation. Furthermore our analysis relies on a crucial assumption regarding the merger time $t_c$. We evaluate $t_c$ as the instant of largest waveform amplitude for the observer orientation with the most prominent double-chirp pattern, i.e. at $\left(\iota_\kappa, \phi_\kappa\right)$. Note that some assumption is necessary here for pure waveform analysis without direct access to the time of common horizon formation as used in the seminal paper \cite{CalderonBustillo2020}.\par

Furthermore, our choice is in contrast to the more conventional assumption in waveform analysis, i.e. the time of the $\left(l,m\right) = \left(2, \pm2\right)$ mode amplitude peak. As discussed in Sec.(\ref{modes}), due to the strong dependence of these waveforms and their time-frequency morphology on higher-order modes, assessing $t_c$ from the $\left(l,m\right) = \left(2, \pm2\right)$ mode alone is insufficient. We can see in Table(\ref{table:cs}) that at the $\left(\iota_\kappa, \phi_\kappa\right)$ orientation the $\left(l,m\right) = \left(2, \pm2\right)$ mode amplitude peak always occurs after that of the full-mode waveform. As such relying on this commonly-used assumption would lead to at best a phase shift in the determination of $\left(\iota_\kappa, \phi_\kappa\right)$, or at worst a misinterpretation of the post-merger behavior.\par

\subsection{Towards inferring the horizon geometry}
\label{future_work}

%The work that we have presented in this document supports the ideas presented in \cite{CalderonBustillo2020}, where for the first time a connection between the observable gravitational waves that we may receive in Earth and the geometry of the post-merger FBH horizon is detailed. This connection offers the tantalizing possibility of imaging black holes directly, and at perhaps the most nuanced moment of their life cycle - the transition from post-merger deformation to spherical symmetry.\par 

The work presented in this paper supports the concepts introduced in \cite{CalderonBustillo2020}, which first established a connection between observable gravitational waves detected on Earth and the geometry of post-merger black hole horizons. This connection suggests the possibility of directly imaging black holes during their most dynamic phase — the transition from post-merger deformation to spherical symmetry. Realizing this possibility requires two parallel efforts: (1) the search for candidate signals and (2) the theoretical framework to extract horizon geometry information.\par 

 The search effort is already  progressing through multiple pipelines designed to detect \gw{} signals with higher-order modal content~\cite{Harry2018, Chandra2022, Sharma2022}. In \cite{CalderonBustillo2020} it is predicted that an optimally oriented $q = 5$ system could be observable by O3 LIGO with a secondary spectral peak signal-to-noise ratio (SNR) of $\rho = 5$ at a distance of $\sim 600\,$Mpc. The figure of merit for \gw{} detector sensitivity is the distance at which a $1.4\, M_\odot - 1.4\, M_\odot$ BNS merger can be observed with SNR of 8~\cite{Chen2020}. At the time of this publication LIGO is in O4, and has reached BNS ranges of $130-165\,$Mpc and $145-177\,$Mpc for the LIGO Hanford and LIGO Livingston detectors respectively~\cite{Capote2024}. It is predicted that these figures could increase to $\sim 330\,$Mpc in O5~\cite{Acernese2018}. The prospect of detecting a candidate double-chirp signal becomes even more feasible with third-generation detectors~\cite{Evans2021, Branchesi2023} that offer a further $10\times$ gain in sensitivity.\par
 
 Our contribution addresses the second effort by elucidating the morphology of the double-chirp pattern and its connection to the underlying physics. We have demonstrated correlations between intrinsic \bbh{} parameters and post-merger spectral peaks revealed through time-frequency mapping. However, definitive statements about horizon geometry require additional numerical relativity (NR) simulations focused on horizon data. These simulations, using codes like \texttt{AHFINDERDIRECT}~\cite{Thornburg2004}, can output spatial data representing horizon surface geometry evolution—information not typically included in waveform catalogs. Note that the exact nature and format of curvature data for dynamical horizons as used in e.g. \cite{CalderonBustillo2020} vary depending on the choice of numerical gauge and coordinate system, and as such consistency is crucial for further study of this phenomenon. This is one of the significant challenges in going forward into the development of methods for inference of horizon geometry - a consistent definition of how horizon geometry data is represented and connected to the underlying physics is needed.\par

Given the establishment of a consistent definition for horizon geometry data, incorporating said data into surrogate waveform models~\cite{Blackman2017} would enable inference of horizon geometry from detected signals, effectively ``imaging" post-merger black holes. Here, ``imaging" refers to statistical inference of markers correlated with horizon geometry. The existence of post-merger spectral peaks indicates such correlation is possible. Past efforts suggest additional levels of correlation between dynamical horizons and observable \gw{} strain~\cite{Rezzolla2010, Owen2011, Nichols2011a, Zhang2012, Nichols2012b, Jaramillo2012, Jaramillo2012a, Jaramillo2012b, Gupta2018, Prasad2020, Mourier2021, Prasad2022, Khera2023}. Our results support the {\it horizon correlation conjecture}~\cite{Prasad2023}, which suggests strong and weak field behaviors are correlated, consistent with the continued trajectory paradigm discussed in Sec.(\ref{evidence}).\par

Finally, the potential yield of statistical information on horizon geometry, its evolution, and correspondence with other parameters would provide a powerful probe of strong-field dynamics in \bbh{} mergers and enable novel tests of general relativity. Surrogate waveform models incorporating geometry data, combined with suitable parameterization of the relevant degrees of freedom and comprehensive understanding of their correlation with intrinsic parameters, could enable horizon geometry inference using methods similar to existing parameter estimation architecture~\cite{lange2018rapid, Ashton_2019, Veitch_2015}.

\section*{Acknowledgements}

We thank Megan Arogeti, Deborah Ferguson, Pablo Laguna, Max Isi, and Vaishak Prasad for helpful discussions and commentary. The authors are grateful for computational resources provided by the LIGO Laboratory. LIGO Laboratory and Advanced LIGO are funded by the United States National Science Foundation (NSF) as well as the Science and Technology Facilities Council (STFC) of the United Kingdom, the Max-Planck-Society (MPS), and the State of Niedersachsen/Germany for support of the construction of Advanced LIGO and construction and operation of the GEO600 detector. Additional support for Advanced LIGO was provided by the Australian Research Council. Virgo is funded, through the European Gravitational Observatory (EGO), by the French Centre National de Recherche Scientifique (CNRS), the Italian Istituto Nazionale di Fisica Nucleare (INFN) and the Dutch Nikhef, with contributions by institutions from Belgium, Germany, Greece, Hungary, Ireland, Japan, Monaco, Poland, Portugal, Spain. KAGRA is supported by the Ministry of Education, Culture, Sports, Science, and Technology (MEXT) in Japan, and is hosted by the Institute for Cosmic Ray Research (ICRR), the University of Tokyo, and co-hosted by High Energy Accelerator Research Organization (KEK) and the National Astronomical Observatory of Japan (NAOJ). This work was supported by NSF grants PHY-2110481 and PHY-2409714.

\appendix
\setcounter{equation}{0}
\renewcommand{\theequation}{A\arabic{equation}}
\section{Peak Finding}
\addcontentsline{toc}{section}{Appendix: Peak Finding}
\label{apndx_peak}

To determine the time location of spectral peaks (see e.g. Fig.(\ref{NR_qvary})), we use the following algorithm. Each time-frequency map is an array of CWT coefficients $T(t, f)$~\cite{Henshaw2024}. We begin by summing all $T$ coefficients per time step over a select frequency band, resulting in a one-dimensional $f_{sum}(t)$ array:

\begin{align}
    f_{sum}(t) = \sum_{f_{floor}}^{f_{high}} T(t, f),
\end{align}

itsem where the lower boundary yof the sum is determined by user input $A$ through $f_{floor} = A*f_{low}$, where $A$ is a scalar. The upper boundary $f_{high}$ is taken as the upper limit of the time-frequency map; this is typically $500\,$Hz. We then look for peaks with the resultant $f_{sum}(t)$ array using the \texttt{scipy.signal.find\_peaks} function~\cite{SciPy}. This function looks for local maxima in a 1D array by comparing each point to its nearest neighbors, and then filtering the results based on user-specified criteria. A given point $x[i]$ in the array is considered a local maximum if $x[i-1] < x[i] > x[i+1]$, where $i$ is the array index of that point. We then use three of the available filters provided by this function: height, prominence, and distance. Each of these is a threshold; any local maxima $x[i]$ that are below each threshold is discarded.\par

The height threshold checks the direct value $x[i]$. This is the broadest filter, and roughly determines the scale of peaks of interest. The prominence of a peak is the difference between that peak's value $x[i]$ and the lowest point in the adjacent valley such that a higher peak is found on that side. If there is no such valley on that side, then the boundary is set as the end of data in that direction. In other words, the prominence is the degree to which a peak stands out relative to nearby adjacent points. This filter allows for very fine tuning of which peaks to retain; it must be set low enough to retain peaks of interest (i.e. physical effects) but high enough to exclude small variations. Distance is the minimum distance between adjacent peaks; this filter can be useful to exclude very close peaks which can occur unintentionally.\par

The exact filter settings and the choice of $A$ (and thus the frequency floor) must be determined on a case-by-case basis; it is very difficult to design a general prescription due to a few different factors. First, $f_{low}$ may be different depending on the waveform; we typically use $f_{low} = 30\,$Hz for the NR waveforms and $f_{low} = 20\,$Hz for the SEOBNRv4PHM waveforms. Setting the frequency floor will determine the height scale of the peaks in the resultant $f_{sum}(t)$ array, as well as their prominence. The floor must be low enough to contain enough power to make the peaks distinguishable, but high enough to avoid the low-frequency content that would bloat the $f_{sum}(t)$ array.\par

Additionally the time segment of the waveform that is being analyzed plays a significant role, as the value of $f_{floor}$ needs to be higher in the merger and post-merger stage to avoid the low-frequency content. This is particularly important when locating the weak additional spectral peaks beyond the nominal double-chirp pattern. As a concrete example, for the $\xi=0.75$ in Fig.(\ref{SEOB_posxivary}) we use $A=5$, height of $0.1$, and prominence of $0.6$ for the inspiral stage, and $A=10$, height of $0.1$, and prominence of $0.05$ for the merger/postmerger stage, with no distance threshold. Notice that in the postmerger stage the height and prominence setting change drastically, as doubling $f_{floor}$ drastically alters the shape of the $f_{sum}(t)$ array.\par

% END PAPER CONTENT

%\centering
%\noindent\rule{8cm}{0.4pt}

%\section*{References \label{refs}}
%\printbibliography[heading=none]
%\clearpage
\bibliography{references}

\begin{thebibliography}{70}%
\makeatletter
\providecommand \@ifxundefined [1]{%
 \@ifx{#1\undefined}
}%
\providecommand \@ifnum [1]{%
 \ifnum #1\expandafter \@firstoftwo
 \else \expandafter \@secondoftwo
 \fi
}%
\providecommand \@ifx [1]{%
 \ifx #1\expandafter \@firstoftwo
 \else \expandafter \@secondoftwo
 \fi
}%
\providecommand \natexlab [1]{#1}%
\providecommand \enquote  [1]{``#1''}%
\providecommand \bibnamefont  [1]{#1}%
\providecommand \bibfnamefont [1]{#1}%
\providecommand \citenamefont [1]{#1}%
\providecommand \href@noop [0]{\@secondoftwo}%
\providecommand \href [0]{\begingroup \@sanitize@url \@href}%
\providecommand \@href[1]{\@@startlink{#1}\@@href}%
\providecommand \@@href[1]{\endgroup#1\@@endlink}%
\providecommand \@sanitize@url [0]{\catcode `\\12\catcode `\$12\catcode `\&12\catcode `\#12\catcode `\^12\catcode `\_12\catcode `\%12\relax}%
\providecommand \@@startlink[1]{}%
\providecommand \@@endlink[0]{}%
\providecommand \url  [0]{\begingroup\@sanitize@url \@url }%
\providecommand \@url [1]{\endgroup\@href {#1}{\urlprefix }}%
\providecommand \urlprefix  [0]{URL }%
\providecommand \Eprint [0]{\href }%
\providecommand \doibase [0]{https://doi.org/}%
\providecommand \selectlanguage [0]{\@gobble}%
\providecommand \bibinfo  [0]{\@secondoftwo}%
\providecommand \bibfield  [0]{\@secondoftwo}%
\providecommand \translation [1]{[#1]}%
\providecommand \BibitemOpen [0]{}%
\providecommand \bibitemStop [0]{}%
\providecommand \bibitemNoStop [0]{.\EOS\space}%
\providecommand \EOS [0]{\spacefactor3000\relax}%
\providecommand \BibitemShut  [1]{\csname bibitem#1\endcsname}%
\let\auto@bib@innerbib\@empty
%</preamble>
\bibitem [{\citenamefont {Abbott}\ \emph {et~al.}(2019)\citenamefont {Abbott} \emph {et~al.}}]{gwtc1}%
  \BibitemOpen
  \bibfield  {author} {\bibinfo {author} {\bibfnamefont {B.~P.}\ \bibnamefont {Abbott}} \emph {et~al.} (\bibinfo {collaboration} {LIGO Scientific, Virgo}),\ }\href {https://doi.org/10.1103/PhysRevX.9.031040} {\bibfield  {journal} {\bibinfo  {journal} {Phys. Rev. X}\ }\textbf {\bibinfo {volume} {9}},\ \bibinfo {pages} {031040} (\bibinfo {year} {2019})},\ \Eprint {https://arxiv.org/abs/1811.12907} {arXiv:1811.12907 [astro-ph.HE]} \BibitemShut {NoStop}%
\bibitem [{\citenamefont {Abbott}\ \emph {et~al.}(2021{\natexlab{a}})\citenamefont {Abbott} \emph {et~al.}}]{gwtc2}%
  \BibitemOpen
  \bibfield  {author} {\bibinfo {author} {\bibfnamefont {R.}~\bibnamefont {Abbott}} \emph {et~al.} (\bibinfo {collaboration} {LIGO Scientific, Virgo}),\ }\href {https://doi.org/10.1103/PhysRevX.11.021053} {\bibfield  {journal} {\bibinfo  {journal} {Phys. Rev. X}\ }\textbf {\bibinfo {volume} {11}},\ \bibinfo {pages} {021053} (\bibinfo {year} {2021}{\natexlab{a}})},\ \Eprint {https://arxiv.org/abs/2010.14527} {arXiv:2010.14527 [gr-qc]} \BibitemShut {NoStop}%
\bibitem [{\citenamefont {Abbott}\ \emph {et~al.}(2024)\citenamefont {Abbott} \emph {et~al.}}]{gwtc2.1}%
  \BibitemOpen
  \bibfield  {author} {\bibinfo {author} {\bibfnamefont {R.}~\bibnamefont {Abbott}} \emph {et~al.} (\bibinfo {collaboration} {LIGO Scientific, VIRGO}),\ }\href {https://doi.org/10.1103/PhysRevD.109.022001} {\bibfield  {journal} {\bibinfo  {journal} {Phys. Rev. D}\ }\textbf {\bibinfo {volume} {109}},\ \bibinfo {pages} {022001} (\bibinfo {year} {2024})},\ \Eprint {https://arxiv.org/abs/2108.01045} {arXiv:2108.01045 [gr-qc]} \BibitemShut {NoStop}%
\bibitem [{\citenamefont {Abbott}\ \emph {et~al.}(2023)\citenamefont {Abbott} \emph {et~al.}}]{gwtc3}%
  \BibitemOpen
  \bibfield  {author} {\bibinfo {author} {\bibfnamefont {R.}~\bibnamefont {Abbott}} \emph {et~al.} (\bibinfo {collaboration} {KAGRA, VIRGO, LIGO Scientific}),\ }\href {https://doi.org/10.1103/PhysRevX.13.041039} {\bibfield  {journal} {\bibinfo  {journal} {Phys. Rev. X}\ }\textbf {\bibinfo {volume} {13}},\ \bibinfo {pages} {041039} (\bibinfo {year} {2023})},\ \Eprint {https://arxiv.org/abs/2111.03606} {arXiv:2111.03606 [gr-qc]} \BibitemShut {NoStop}%
\bibitem [{\citenamefont {Abbott}\ \emph {et~al.}(2017)\citenamefont {Abbott} \emph {et~al.}}]{GW170817}%
  \BibitemOpen
  \bibfield  {author} {\bibinfo {author} {\bibfnamefont {B.~P.}\ \bibnamefont {Abbott}} \emph {et~al.} (\bibinfo {collaboration} {LIGO Scientific, Virgo}),\ }\href {https://doi.org/10.1103/PhysRevLett.119.161101} {\bibfield  {journal} {\bibinfo  {journal} {Phys. Rev. Lett.}\ }\textbf {\bibinfo {volume} {119}},\ \bibinfo {pages} {161101} (\bibinfo {year} {2017})},\ \Eprint {https://arxiv.org/abs/1710.05832} {arXiv:1710.05832 [gr-qc]} \BibitemShut {NoStop}%
\bibitem [{\citenamefont {Abbott}\ \emph {et~al.}(2020{\natexlab{a}})\citenamefont {Abbott} \emph {et~al.}}]{GW190425}%
  \BibitemOpen
  \bibfield  {author} {\bibinfo {author} {\bibfnamefont {B.~P.}\ \bibnamefont {Abbott}} \emph {et~al.} (\bibinfo {collaboration} {LIGO Scientific, Virgo}),\ }\href {https://doi.org/10.3847/2041-8213/ab75f5} {\bibfield  {journal} {\bibinfo  {journal} {Astrophys. J. Lett.}\ }\textbf {\bibinfo {volume} {892}},\ \bibinfo {pages} {L3} (\bibinfo {year} {2020}{\natexlab{a}})},\ \Eprint {https://arxiv.org/abs/2001.01761} {arXiv:2001.01761 [astro-ph.HE]} \BibitemShut {NoStop}%
\bibitem [{\citenamefont {Abbott}\ \emph {et~al.}(2020{\natexlab{b}})\citenamefont {Abbott} \emph {et~al.}}]{GW190814}%
  \BibitemOpen
  \bibfield  {author} {\bibinfo {author} {\bibfnamefont {R.}~\bibnamefont {Abbott}} \emph {et~al.} (\bibinfo {collaboration} {LIGO Scientific, Virgo}),\ }\href {https://doi.org/10.3847/2041-8213/ab960f} {\bibfield  {journal} {\bibinfo  {journal} {Astrophys. J. Lett.}\ }\textbf {\bibinfo {volume} {896}},\ \bibinfo {pages} {L44} (\bibinfo {year} {2020}{\natexlab{b}})},\ \Eprint {https://arxiv.org/abs/2006.12611} {arXiv:2006.12611 [astro-ph.HE]} \BibitemShut {NoStop}%
\bibitem [{\citenamefont {Abbott}\ \emph {et~al.}(2021{\natexlab{b}})\citenamefont {Abbott} \emph {et~al.}}]{NSBHx2}%
  \BibitemOpen
  \bibfield  {author} {\bibinfo {author} {\bibfnamefont {R.}~\bibnamefont {Abbott}} \emph {et~al.} (\bibinfo {collaboration} {LIGO Scientific, KAGRA, VIRGO}),\ }\href {https://doi.org/10.3847/2041-8213/ac082e} {\bibfield  {journal} {\bibinfo  {journal} {Astrophys. J. Lett.}\ }\textbf {\bibinfo {volume} {915}},\ \bibinfo {pages} {L5} (\bibinfo {year} {2021}{\natexlab{b}})},\ \Eprint {https://arxiv.org/abs/2106.15163} {arXiv:2106.15163 [astro-ph.HE]} \BibitemShut {NoStop}%
\bibitem [{\citenamefont {Abac}\ \emph {et~al.}(2024)\citenamefont {Abac} \emph {et~al.}}]{O4_NSBH}%
  \BibitemOpen
  \bibfield  {author} {\bibinfo {author} {\bibfnamefont {A.~G.}\ \bibnamefont {Abac}} \emph {et~al.} (\bibinfo {collaboration} {LIGO Scientific, KAGRA, VIRGO}),\ }\href {https://doi.org/10.3847/2041-8213/ad5beb} {\bibfield  {journal} {\bibinfo  {journal} {Astrophys. J. Lett.}\ }\textbf {\bibinfo {volume} {970}},\ \bibinfo {pages} {L34} (\bibinfo {year} {2024})},\ \Eprint {https://arxiv.org/abs/2404.04248} {arXiv:2404.04248 [astro-ph.HE]} \BibitemShut {NoStop}%
\bibitem [{\citenamefont {Lincoln}\ and\ \citenamefont {Will}(1990)}]{Lincoln1990}%
  \BibitemOpen
  \bibfield  {author} {\bibinfo {author} {\bibfnamefont {C.~W.}\ \bibnamefont {Lincoln}}\ and\ \bibinfo {author} {\bibfnamefont {C.~M.}\ \bibnamefont {Will}},\ }\href {https://doi.org/10.1103/PhysRevD.42.1123} {\bibfield  {journal} {\bibinfo  {journal} {Physical Review D}\ }\textbf {\bibinfo {volume} {42}},\ \bibinfo {pages} {1123} (\bibinfo {year} {1990})}\BibitemShut {NoStop}%
\bibitem [{\citenamefont {Kidder}(1995)}]{PhysRevD.52.821}%
  \BibitemOpen
  \bibfield  {author} {\bibinfo {author} {\bibfnamefont {L.~E.}\ \bibnamefont {Kidder}},\ }\href {https://doi.org/10.1103/PhysRevD.52.821} {\bibfield  {journal} {\bibinfo  {journal} {Phys. Rev. D}\ }\textbf {\bibinfo {volume} {52}},\ \bibinfo {pages} {821} (\bibinfo {year} {1995})}\BibitemShut {NoStop}%
\bibitem [{\citenamefont {Racine}(2008)}]{Racine_2008}%
  \BibitemOpen
  \bibfield  {author} {\bibinfo {author} {\bibfnamefont {E.}~\bibnamefont {Racine}},\ }\bibfield  {journal} {\bibinfo  {journal} {Physical Review D}\ }\textbf {\bibinfo {volume} {78}},\ \href {https://doi.org/10.1103/physrevd.78.044021} {10.1103/physrevd.78.044021} (\bibinfo {year} {2008})\BibitemShut {NoStop}%
\bibitem [{\citenamefont {Read}\ \emph {et~al.}(2009)\citenamefont {Read} \emph {et~al.}}]{Read2009}%
  \BibitemOpen
  \bibfield  {author} {\bibinfo {author} {\bibfnamefont {J.~S.}\ \bibnamefont {Read}} \emph {et~al.},\ }\href {https://doi.org/10.1103/PhysRevD.79.124033} {\bibfield  {journal} {\bibinfo  {journal} {Physical Review D - Particles, Fields, Gravitation and Cosmology}\ }\textbf {\bibinfo {volume} {79}},\ \bibinfo {pages} {1} (\bibinfo {year} {2009})},\ \Eprint {https://arxiv.org/abs/0901.3258} {arXiv:0901.3258} \BibitemShut {NoStop}%
\bibitem [{\citenamefont {Blanchet}(2014)}]{Blanchet2014}%
  \BibitemOpen
  \bibfield  {author} {\bibinfo {author} {\bibfnamefont {L.}~\bibnamefont {Blanchet}},\ }\bibfield  {journal} {\bibinfo  {journal} {Living Reviews in Relativity}\ }\textbf {\bibinfo {volume} {17}},\ \href {https://doi.org/10.12942/lrr-2014-2} {10.12942/lrr-2014-2} (\bibinfo {year} {2014}),\ \Eprint {https://arxiv.org/abs/1310.1528} {arXiv:1310.1528} \BibitemShut {NoStop}%
\bibitem [{\citenamefont {Clark}\ \emph {et~al.}(2016)\citenamefont {Clark}, \citenamefont {Bauswein}, \citenamefont {Stergioulas},\ and\ \citenamefont {Shoemaker}}]{Clark2016}%
  \BibitemOpen
  \bibfield  {author} {\bibinfo {author} {\bibfnamefont {J.~A.}\ \bibnamefont {Clark}}, \bibinfo {author} {\bibfnamefont {A.}~\bibnamefont {Bauswein}}, \bibinfo {author} {\bibfnamefont {N.}~\bibnamefont {Stergioulas}},\ and\ \bibinfo {author} {\bibfnamefont {D.}~\bibnamefont {Shoemaker}},\ }\bibfield  {journal} {\bibinfo  {journal} {Classical and Quantum Gravity}\ }\textbf {\bibinfo {volume} {33}},\ \href {https://doi.org/10.1088/0264-9381/33/8/085003} {10.1088/0264-9381/33/8/085003} (\bibinfo {year} {2016}),\ \Eprint {https://arxiv.org/abs/1509.08522} {arXiv:1509.08522} \BibitemShut {NoStop}%
\bibitem [{\citenamefont {Bauswein}\ \emph {et~al.}(2016)\citenamefont {Bauswein}, \citenamefont {Stergioulas},\ and\ \citenamefont {Janka}}]{Bauswein2016}%
  \BibitemOpen
  \bibfield  {author} {\bibinfo {author} {\bibfnamefont {A.}~\bibnamefont {Bauswein}}, \bibinfo {author} {\bibfnamefont {N.}~\bibnamefont {Stergioulas}},\ and\ \bibinfo {author} {\bibfnamefont {H.~T.}\ \bibnamefont {Janka}},\ }\bibfield  {journal} {\bibinfo  {journal} {European Physical Journal A}\ }\textbf {\bibinfo {volume} {52}},\ \href {https://doi.org/10.1140/epja/i2016-16056-7} {10.1140/epja/i2016-16056-7} (\bibinfo {year} {2016}),\ \Eprint {https://arxiv.org/abs/1508.05493} {arXiv:1508.05493} \BibitemShut {NoStop}%
\bibitem [{\citenamefont {Schmidt}\ \emph {et~al.}(2015)\citenamefont {Schmidt}, \citenamefont {Ohme},\ and\ \citenamefont {Hannam}}]{Schmidt_2015}%
  \BibitemOpen
  \bibfield  {author} {\bibinfo {author} {\bibfnamefont {P.}~\bibnamefont {Schmidt}}, \bibinfo {author} {\bibfnamefont {F.}~\bibnamefont {Ohme}},\ and\ \bibinfo {author} {\bibfnamefont {M.}~\bibnamefont {Hannam}},\ }\bibfield  {journal} {\bibinfo  {journal} {Physical Review D}\ }\textbf {\bibinfo {volume} {91}},\ \href {https://doi.org/10.1103/physrevd.91.024043} {10.1103/physrevd.91.024043} (\bibinfo {year} {2015})\BibitemShut {NoStop}%
\bibitem [{\citenamefont {Gerosa}\ \emph {et~al.}(2021)\citenamefont {Gerosa} \emph {et~al.}}]{Gerosa_2021}%
  \BibitemOpen
  \bibfield  {author} {\bibinfo {author} {\bibfnamefont {D.}~\bibnamefont {Gerosa}} \emph {et~al.},\ }\bibfield  {journal} {\bibinfo  {journal} {Physical Review D}\ }\textbf {\bibinfo {volume} {103}},\ \href {https://doi.org/10.1103/physrevd.103.064067} {10.1103/physrevd.103.064067} (\bibinfo {year} {2021})\BibitemShut {NoStop}%
\bibitem [{\citenamefont {Henshaw}\ \emph {et~al.}(2022)\citenamefont {Henshaw}, \citenamefont {O'Shaughnessy},\ and\ \citenamefont {Cadonati}}]{Henshaw2022}%
  \BibitemOpen
  \bibfield  {author} {\bibinfo {author} {\bibfnamefont {C.}~\bibnamefont {Henshaw}}, \bibinfo {author} {\bibfnamefont {R.}~\bibnamefont {O'Shaughnessy}},\ and\ \bibinfo {author} {\bibfnamefont {L.}~\bibnamefont {Cadonati}},\ }\href {https://doi.org/10.1088/1361-6382/ac6cc0} {\bibfield  {journal} {\bibinfo  {journal} {Classical and Quantum Gravity}\ }\textbf {\bibinfo {volume} {39}},\ \bibinfo {pages} {125003} (\bibinfo {year} {2022})},\ \Eprint {https://arxiv.org/abs/2201.05220} {arXiv:2201.05220} \BibitemShut {NoStop}%
\bibitem [{\citenamefont {Gerosa}\ \emph {et~al.}(2023)\citenamefont {Gerosa} \emph {et~al.}}]{Gerosa2023}%
  \BibitemOpen
  \bibfield  {author} {\bibinfo {author} {\bibfnamefont {D.}~\bibnamefont {Gerosa}} \emph {et~al.},\ }\href {https://doi.org/10.1103/PhysRevD.108.024042} {\bibfield  {journal} {\bibinfo  {journal} {Phys. Rev. D}\ }\textbf {\bibinfo {volume} {108}},\ \bibinfo {pages} {024042} (\bibinfo {year} {2023})}\BibitemShut {NoStop}%
\bibitem [{\citenamefont {Antonini}\ and\ \citenamefont {Gieles}(2020)}]{Antonini_2020}%
  \BibitemOpen
  \bibfield  {author} {\bibinfo {author} {\bibfnamefont {F.}~\bibnamefont {Antonini}}\ and\ \bibinfo {author} {\bibfnamefont {M.}~\bibnamefont {Gieles}},\ }\bibfield  {journal} {\bibinfo  {journal} {Physical Review D}\ }\textbf {\bibinfo {volume} {102}},\ \href {https://doi.org/10.1103/physrevd.102.123016} {10.1103/physrevd.102.123016} (\bibinfo {year} {2020})\BibitemShut {NoStop}%
\bibitem [{\citenamefont {Trani}\ \emph {et~al.}(2021)\citenamefont {Trani} \emph {et~al.}}]{Trani_2021}%
  \BibitemOpen
  \bibfield  {author} {\bibinfo {author} {\bibfnamefont {A.~A.}\ \bibnamefont {Trani}} \emph {et~al.},\ }\href {https://doi.org/10.1093/mnras/stab967} {\bibfield  {journal} {\bibinfo  {journal} {Monthly Notices of the Royal Astronomical Society}\ }\textbf {\bibinfo {volume} {504}},\ \bibinfo {pages} {910–919} (\bibinfo {year} {2021})}\BibitemShut {NoStop}%
\bibitem [{\citenamefont {Belczynski}\ \emph {et~al.}(2020)\citenamefont {Belczynski} \emph {et~al.}}]{Belczynski_2020}%
  \BibitemOpen
  \bibfield  {author} {\bibinfo {author} {\bibfnamefont {K.}~\bibnamefont {Belczynski}} \emph {et~al.},\ }\href {https://doi.org/10.1051/0004-6361/201936528} {\bibfield  {journal} {\bibinfo  {journal} {Astronomy \& Astrophysics}\ }\textbf {\bibinfo {volume} {636}},\ \bibinfo {pages} {A104} (\bibinfo {year} {2020})}\BibitemShut {NoStop}%
\bibitem [{\citenamefont {{Calderon Bustillo}}\ \emph {et~al.}(2020)\citenamefont {{Calderon Bustillo}} \emph {et~al.}}]{CalderonBustillo2020}%
  \BibitemOpen
  \bibfield  {author} {\bibinfo {author} {\bibfnamefont {J.}~\bibnamefont {{Calderon Bustillo}}} \emph {et~al.},\ }\href {https://doi.org/10.1038/s42005-020-00446-7} {\bibfield  {journal} {\bibinfo  {journal} {Communications Physics}\ }\textbf {\bibinfo {volume} {3}},\ \bibinfo {pages} {1} (\bibinfo {year} {2020})},\ \Eprint {https://arxiv.org/abs/1906.01153} {arXiv:1906.01153} \BibitemShut {NoStop}%
\bibitem [{\citenamefont {Harry}\ \emph {et~al.}(2018)\citenamefont {Harry}, \citenamefont {Bustillo},\ and\ \citenamefont {Nitz}}]{Harry2018}%
  \BibitemOpen
  \bibfield  {author} {\bibinfo {author} {\bibfnamefont {I.}~\bibnamefont {Harry}}, \bibinfo {author} {\bibfnamefont {J.~C.}\ \bibnamefont {Bustillo}},\ and\ \bibinfo {author} {\bibfnamefont {A.}~\bibnamefont {Nitz}},\ }\href {https://doi.org/10.1103/PhysRevD.97.023004} {\bibfield  {journal} {\bibinfo  {journal} {Physical Review D}\ }\textbf {\bibinfo {volume} {97}},\ \bibinfo {pages} {23004} (\bibinfo {year} {2018})},\ \Eprint {https://arxiv.org/abs/1709.09181} {arXiv:1709.09181} \BibitemShut {NoStop}%
\bibitem [{\citenamefont {Chandra}\ \emph {et~al.}(2022)\citenamefont {Chandra}, \citenamefont {Calder\'on~Bustillo}, \citenamefont {Pai},\ and\ \citenamefont {Harry}}]{Chandra2022}%
  \BibitemOpen
  \bibfield  {author} {\bibinfo {author} {\bibfnamefont {K.}~\bibnamefont {Chandra}}, \bibinfo {author} {\bibfnamefont {J.}~\bibnamefont {Calder\'on~Bustillo}}, \bibinfo {author} {\bibfnamefont {A.}~\bibnamefont {Pai}},\ and\ \bibinfo {author} {\bibfnamefont {I.~W.}\ \bibnamefont {Harry}},\ }\href {https://doi.org/10.1103/PhysRevD.106.123003} {\bibfield  {journal} {\bibinfo  {journal} {Phys. Rev. D}\ }\textbf {\bibinfo {volume} {106}},\ \bibinfo {pages} {123003} (\bibinfo {year} {2022})},\ \Eprint {https://arxiv.org/abs/2207.01654} {arXiv:2207.01654 [gr-qc]} \BibitemShut {NoStop}%
\bibitem [{\citenamefont {Sharma}\ \emph {et~al.}(2022)\citenamefont {Sharma}, \citenamefont {Chandra},\ and\ \citenamefont {Pai}}]{Sharma2022}%
  \BibitemOpen
  \bibfield  {author} {\bibinfo {author} {\bibfnamefont {K.}~\bibnamefont {Sharma}}, \bibinfo {author} {\bibfnamefont {K.}~\bibnamefont {Chandra}},\ and\ \bibinfo {author} {\bibfnamefont {A.}~\bibnamefont {Pai}},\ }\href {http://arxiv.org/abs/2208.02545} {} (\bibinfo {year} {2022}),\ \Eprint {https://arxiv.org/abs/2208.02545} {arXiv:2208.02545} \BibitemShut {NoStop}%
\bibitem [{\citenamefont {Penrose}(1965)}]{Penrose1965}%
  \BibitemOpen
  \bibfield  {author} {\bibinfo {author} {\bibfnamefont {R.}~\bibnamefont {Penrose}},\ }\href {https://doi.org/10.1103/PhysRevLett.14.57} {\bibfield  {journal} {\bibinfo  {journal} {Physical Review Letters}\ }\textbf {\bibinfo {volume} {14}},\ \bibinfo {pages} {57} (\bibinfo {year} {1965})}\BibitemShut {NoStop}%
\bibitem [{\citenamefont {Ashtekar}\ and\ \citenamefont {Krishnan}(2004)}]{Ashtekar2004}%
  \BibitemOpen
  \bibfield  {author} {\bibinfo {author} {\bibfnamefont {A.}~\bibnamefont {Ashtekar}}\ and\ \bibinfo {author} {\bibfnamefont {B.}~\bibnamefont {Krishnan}},\ }\bibfield  {journal} {\bibinfo  {journal} {Living Reviews in Relativity}\ }\textbf {\bibinfo {volume} {7}},\ \href {https://doi.org/10.12942/lrr-2004-10} {10.12942/lrr-2004-10} (\bibinfo {year} {2004}),\ \Eprint {https://arxiv.org/abs/0407042} {arXiv:0407042 [gr-qc]} \BibitemShut {NoStop}%
\bibitem [{\citenamefont {Booth}(2005)}]{Booth2005}%
  \BibitemOpen
  \bibfield  {author} {\bibinfo {author} {\bibfnamefont {I.}~\bibnamefont {Booth}},\ }\href {https://doi.org/10.1139/p05-063} {\bibfield  {journal} {\bibinfo  {journal} {Canadian Journal of Physics}\ }\textbf {\bibinfo {volume} {83}},\ \bibinfo {pages} {1073} (\bibinfo {year} {2005})},\ \Eprint {https://arxiv.org/abs/0508107} {arXiv:0508107 [gr-qc]} \BibitemShut {NoStop}%
\bibitem [{\citenamefont {Boyle}\ \emph {et~al.}(2019)\citenamefont {Boyle} \emph {et~al.}}]{SXS2019}%
  \BibitemOpen
  \bibfield  {author} {\bibinfo {author} {\bibfnamefont {M.}~\bibnamefont {Boyle}} \emph {et~al.},\ }\bibfield  {journal} {\bibinfo  {journal} {Classical and Quantum Gravity}\ }\textbf {\bibinfo {volume} {36}},\ \href {https://doi.org/10.1088/1361-6382/ab34e2} {10.1088/1361-6382/ab34e2} (\bibinfo {year} {2019}),\ \Eprint {https://arxiv.org/abs/1904.04831} {arXiv:1904.04831} \BibitemShut {NoStop}%
\bibitem [{\citenamefont {Jani}\ \emph {et~al.}(2016)\citenamefont {Jani} \emph {et~al.}}]{Jani2016}%
  \BibitemOpen
  \bibfield  {author} {\bibinfo {author} {\bibfnamefont {K.}~\bibnamefont {Jani}} \emph {et~al.},\ }\href {https://doi.org/10.1088/0264-9381/33/20/204001} {\bibfield  {journal} {\bibinfo  {journal} {Classical and Quantum Gravity}\ }\textbf {\bibinfo {volume} {33}},\ \bibinfo {pages} {1} (\bibinfo {year} {2016})},\ \Eprint {https://arxiv.org/abs/1605.03204} {arXiv:1605.03204} \BibitemShut {NoStop}%
\bibitem [{\citenamefont {Ferguson}\ \emph {et~al.}(2023)\citenamefont {Ferguson} \emph {et~al.}}]{Ferguson2023}%
  \BibitemOpen
  \bibfield  {author} {\bibinfo {author} {\bibfnamefont {D.}~\bibnamefont {Ferguson}} \emph {et~al.},\ }\href {http://arxiv.org/abs/2309.00262} {} (\bibinfo {year} {2023}),\ \Eprint {https://arxiv.org/abs/2309.00262} {arXiv:2309.00262} \BibitemShut {NoStop}%
\bibitem [{\citenamefont {Healy}\ and\ \citenamefont {Lousto}(2022)}]{Healy2022}%
  \BibitemOpen
  \bibfield  {author} {\bibinfo {author} {\bibfnamefont {J.}~\bibnamefont {Healy}}\ and\ \bibinfo {author} {\bibfnamefont {C.~O.}\ \bibnamefont {Lousto}},\ }\bibfield  {journal} {\bibinfo  {journal} {Physical Review D}\ }\textbf {\bibinfo {volume} {105}},\ \href {https://doi.org/10.1103/PhysRevD.105.124010} {10.1103/PhysRevD.105.124010} (\bibinfo {year} {2022}),\ \Eprint {https://arxiv.org/abs/2202.00018} {arXiv:2202.00018} \BibitemShut {NoStop}%
\bibitem [{\citenamefont {Nitz}\ \emph {et~al.}(2023)\citenamefont {Nitz} \emph {et~al.}}]{pycbc}%
  \BibitemOpen
  \bibfield  {author} {\bibinfo {author} {\bibfnamefont {A.}~\bibnamefont {Nitz}} \emph {et~al.},\ }\href {https://doi.org/10.5281/zenodo.7692098} {\bibinfo {title} {gwastro/pycbc: v2.1.0 release of pycbc}} (\bibinfo {year} {2023})\BibitemShut {NoStop}%
\bibitem [{\citenamefont {Ossokine}\ \emph {et~al.}(2020)\citenamefont {Ossokine} \emph {et~al.}}]{SEOBNRv4PHM}%
  \BibitemOpen
  \bibfield  {author} {\bibinfo {author} {\bibfnamefont {S.}~\bibnamefont {Ossokine}} \emph {et~al.},\ }\href {https://doi.org/10.1103/PhysRevD.102.044055} {\bibfield  {journal} {\bibinfo  {journal} {Phys. Rev. D}\ }\textbf {\bibinfo {volume} {102}},\ \bibinfo {pages} {044055} (\bibinfo {year} {2020})}\BibitemShut {NoStop}%
\bibitem [{\citenamefont {Henshaw}\ \emph {et~al.}(2024)\citenamefont {Henshaw}, \citenamefont {Arogeti}, \citenamefont {Heranval},\ and\ \citenamefont {Cadonati}}]{Henshaw2024}%
  \BibitemOpen
  \bibfield  {author} {\bibinfo {author} {\bibfnamefont {C.}~\bibnamefont {Henshaw}}, \bibinfo {author} {\bibfnamefont {M.}~\bibnamefont {Arogeti}}, \bibinfo {author} {\bibfnamefont {A.}~\bibnamefont {Heranval}},\ and\ \bibinfo {author} {\bibfnamefont {L.}~\bibnamefont {Cadonati}},\ }\Eprint {https://arxiv.org/abs/2402.16533} {arXiv:2402.16533 [gr-qc]}  (\bibinfo {year} {2024})\BibitemShut {NoStop}%
\bibitem [{\citenamefont {Mehta}\ \emph {et~al.}(2017)\citenamefont {Mehta}, \citenamefont {Mishra}, \citenamefont {Varma},\ and\ \citenamefont {Ajith}}]{Mehta2017}%
  \BibitemOpen
  \bibfield  {author} {\bibinfo {author} {\bibfnamefont {A.~K.}\ \bibnamefont {Mehta}}, \bibinfo {author} {\bibfnamefont {C.~K.}\ \bibnamefont {Mishra}}, \bibinfo {author} {\bibfnamefont {V.}~\bibnamefont {Varma}},\ and\ \bibinfo {author} {\bibfnamefont {P.}~\bibnamefont {Ajith}},\ }\href {https://doi.org/10.1103/PhysRevD.96.124010} {\bibfield  {journal} {\bibinfo  {journal} {Physical Review D}\ }\textbf {\bibinfo {volume} {96}},\ \bibinfo {pages} {1} (\bibinfo {year} {2017})},\ \Eprint {https://arxiv.org/abs/1708.03501} {arXiv:1708.03501} \BibitemShut {NoStop}%
\bibitem [{\citenamefont {Varma}\ \emph {et~al.}(2019)\citenamefont {Varma}, \citenamefont {Gerosa}, \citenamefont {Stein}, \citenamefont {H{\'{e}}bert},\ and\ \citenamefont {Zhang}}]{Varma2019a}%
  \BibitemOpen
  \bibfield  {author} {\bibinfo {author} {\bibfnamefont {V.}~\bibnamefont {Varma}}, \bibinfo {author} {\bibfnamefont {D.}~\bibnamefont {Gerosa}}, \bibinfo {author} {\bibfnamefont {L.~C.}\ \bibnamefont {Stein}}, \bibinfo {author} {\bibfnamefont {F.}~\bibnamefont {H{\'{e}}bert}},\ and\ \bibinfo {author} {\bibfnamefont {H.}~\bibnamefont {Zhang}},\ }\href {https://doi.org/10.1103/PhysRevLett.122.011101} {\bibfield  {journal} {\bibinfo  {journal} {Physical Review Letters}\ }\textbf {\bibinfo {volume} {122}},\ \bibinfo {pages} {1} (\bibinfo {year} {2019})},\ \Eprint {https://arxiv.org/abs/1809.09125} {arXiv:1809.09125} \BibitemShut {NoStop}%
\bibitem [{\citenamefont {Newman}\ and\ \citenamefont {Penrose}(1962)}]{Newman1962}%
  \BibitemOpen
  \bibfield  {author} {\bibinfo {author} {\bibfnamefont {E.}~\bibnamefont {Newman}}\ and\ \bibinfo {author} {\bibfnamefont {R.}~\bibnamefont {Penrose}},\ }\href {https://doi.org/10.1063/1.1724257} {\bibfield  {journal} {\bibinfo  {journal} {Journal of Mathematical Physics}\ }\textbf {\bibinfo {volume} {3}},\ \bibinfo {pages} {566} (\bibinfo {year} {1962})}\BibitemShut {NoStop}%
\bibitem [{\citenamefont {Pook-kolb}\ \emph {et~al.}(2021)\citenamefont {Pook-kolb}, \citenamefont {Hennigar},\ and\ \citenamefont {Booth}}]{Pook-Kolb2021a}%
  \BibitemOpen
  \bibfield  {author} {\bibinfo {author} {\bibfnamefont {D.}~\bibnamefont {Pook-kolb}}, \bibinfo {author} {\bibfnamefont {R.~A.}\ \bibnamefont {Hennigar}},\ and\ \bibinfo {author} {\bibfnamefont {I.}~\bibnamefont {Booth}},\ }\href {https://doi.org/10.1103/PhysRevLett.127.181101} {\bibfield  {journal} {\bibinfo  {journal} {Physical Review Letters}\ }\textbf {\bibinfo {volume} {127}},\ \bibinfo {pages} {181101} (\bibinfo {year} {2021})}\BibitemShut {NoStop}%
\bibitem [{\citenamefont {Pook-Kolb}\ \emph {et~al.}(2021)\citenamefont {Pook-Kolb}, \citenamefont {Booth},\ and\ \citenamefont {Hennigar}}]{Pook-Kolb2021b}%
  \BibitemOpen
  \bibfield  {author} {\bibinfo {author} {\bibfnamefont {D.}~\bibnamefont {Pook-Kolb}}, \bibinfo {author} {\bibfnamefont {I.}~\bibnamefont {Booth}},\ and\ \bibinfo {author} {\bibfnamefont {R.~A.}\ \bibnamefont {Hennigar}},\ }\href {https://doi.org/10.1103/PhysRevD.104.084084} {\bibfield  {journal} {\bibinfo  {journal} {Physical Review D}\ }\textbf {\bibinfo {volume} {104}},\ \bibinfo {pages} {84084} (\bibinfo {year} {2021})},\ \Eprint {https://arxiv.org/abs/2104.11344} {arXiv:2104.11344} \BibitemShut {NoStop}%
\bibitem [{\citenamefont {Evans}\ \emph {et~al.}(2020)\citenamefont {Evans} \emph {et~al.}}]{Evans2020}%
  \BibitemOpen
  \bibfield  {author} {\bibinfo {author} {\bibfnamefont {C.}~\bibnamefont {Evans}} \emph {et~al.},\ }\bibfield  {journal} {\bibinfo  {journal} {Classical and Quantum Gravity}\ }\textbf {\bibinfo {volume} {37}},\ \href {https://doi.org/10.1088/1361-6382/ab9c6b} {10.1088/1361-6382/ab9c6b} (\bibinfo {year} {2020})\BibitemShut {NoStop}%
\bibitem [{\citenamefont {Booth}\ \emph {et~al.}(2020)\citenamefont {Booth}, \citenamefont {Hennigar},\ and\ \citenamefont {Mondal}}]{Booth2020}%
  \BibitemOpen
  \bibfield  {author} {\bibinfo {author} {\bibfnamefont {I.}~\bibnamefont {Booth}}, \bibinfo {author} {\bibfnamefont {R.~A.}\ \bibnamefont {Hennigar}},\ and\ \bibinfo {author} {\bibfnamefont {S.}~\bibnamefont {Mondal}},\ }\href {https://doi.org/10.1103/PhysRevD.102.044031} {\bibfield  {journal} {\bibinfo  {journal} {Physical Review D}\ }\textbf {\bibinfo {volume} {102}},\ \bibinfo {pages} {44031} (\bibinfo {year} {2020})}\BibitemShut {NoStop}%
\bibitem [{\citenamefont {Schmidt}\ \emph {et~al.}(2012)\citenamefont {Schmidt}, \citenamefont {Hannam},\ and\ \citenamefont {Husa}}]{Schmidt2012}%
  \BibitemOpen
  \bibfield  {author} {\bibinfo {author} {\bibfnamefont {P.}~\bibnamefont {Schmidt}}, \bibinfo {author} {\bibfnamefont {M.}~\bibnamefont {Hannam}},\ and\ \bibinfo {author} {\bibfnamefont {S.}~\bibnamefont {Husa}},\ }\href {https://doi.org/10.1103/PhysRevD.86.104063} {\bibfield  {journal} {\bibinfo  {journal} {Physical Review D - Particles, Fields, Gravitation and Cosmology}\ }\textbf {\bibinfo {volume} {86}},\ \bibinfo {pages} {1} (\bibinfo {year} {2012})},\ \Eprint {https://arxiv.org/abs/1207.3088} {arXiv:1207.3088} \BibitemShut {NoStop}%
\bibitem [{\citenamefont {Chen}\ \emph {et~al.}(2020)\citenamefont {Chen} \emph {et~al.}}]{Chen2020}%
  \BibitemOpen
  \bibfield  {author} {\bibinfo {author} {\bibfnamefont {H.~Y.}\ \bibnamefont {Chen}} \emph {et~al.},\ }\bibfield  {journal} {\bibinfo  {journal} {Classical and Quantum Gravity}\ }\textbf {\bibinfo {volume} {38}},\ \href {https://doi.org/10.1088/1361-6382/abd594} {10.1088/1361-6382/abd594} (\bibinfo {year} {2020}),\ \Eprint {https://arxiv.org/abs/1709.08079} {arXiv:1709.08079} \BibitemShut {NoStop}%
\bibitem [{\citenamefont {Capote}\ \emph {et~al.}(2024)\citenamefont {Capote} \emph {et~al.}}]{Capote2024}%
  \BibitemOpen
  \bibfield  {author} {\bibinfo {author} {\bibfnamefont {E.}~\bibnamefont {Capote}} \emph {et~al.},\ }\href {https://doi.org/10.1103/PhysRevD.111.062002} {\bibfield  {journal} {\bibinfo  {journal} {Physical Review D}\ }\textbf {\bibinfo {volume} {111}},\ \bibinfo {pages} {62002} (\bibinfo {year} {2024})},\ \Eprint {https://arxiv.org/abs/2411.14607} {arXiv:2411.14607} \BibitemShut {NoStop}%
\bibitem [{\citenamefont {Acernese}\ \emph {et~al.}(2018)\citenamefont {Acernese} \emph {et~al.}}]{Acernese2018}%
  \BibitemOpen
  \bibfield  {author} {\bibinfo {author} {\bibfnamefont {F.}~\bibnamefont {Acernese}} \emph {et~al.},\ }\href@noop {} {\bibfield  {journal} {\bibinfo  {journal} {Living Reviews in Relativity}\ }\textbf {\bibinfo {volume} {21}} (\bibinfo {year} {2018})}\BibitemShut {NoStop}%
\bibitem [{\citenamefont {Evans}\ \emph {et~al.}(2021)\citenamefont {Evans} \emph {et~al.}}]{Evans2021}%
  \BibitemOpen
  \bibfield  {author} {\bibinfo {author} {\bibfnamefont {M.}~\bibnamefont {Evans}} \emph {et~al.},\ }\Eprint {https://arxiv.org/abs/2109.09882} {arXiv:2109.09882}  (\bibinfo {year} {2021})\BibitemShut {NoStop}%
\bibitem [{\citenamefont {Branchesi}\ \emph {et~al.}(2023)\citenamefont {Branchesi} \emph {et~al.}}]{Branchesi2023}%
  \BibitemOpen
  \bibfield  {author} {\bibinfo {author} {\bibfnamefont {M.}~\bibnamefont {Branchesi}} \emph {et~al.},\ }\href {https://doi.org/10.1088/1475-7516/2023/07/068} {\bibfield  {journal} {\bibinfo  {journal} {Journal of Cosmology and Astroparticle Physics}\ }\textbf {\bibinfo {volume} {2023}}\bibfield  {number} {\bibinfo  {number} { (7)}},\ }\Eprint {https://arxiv.org/abs/2303.15923} {arXiv:2303.15923} \BibitemShut {NoStop}%
\bibitem [{\citenamefont {Thornburg}(2004)}]{Thornburg2004}%
  \BibitemOpen
  \bibfield  {author} {\bibinfo {author} {\bibfnamefont {J.}~\bibnamefont {Thornburg}},\ }\href {https://doi.org/10.1088/0264-9381/21/2/026} {\bibfield  {journal} {\bibinfo  {journal} {Classical and Quantum Gravity}\ }\textbf {\bibinfo {volume} {21}},\ \bibinfo {pages} {743} (\bibinfo {year} {2004})},\ \Eprint {https://arxiv.org/abs/0306056} {arXiv:0306056 [gr-qc]} \BibitemShut {NoStop}%
\bibitem [{\citenamefont {Blackman}\ \emph {et~al.}(2017)\citenamefont {Blackman} \emph {et~al.}}]{Blackman2017}%
  \BibitemOpen
  \bibfield  {author} {\bibinfo {author} {\bibfnamefont {J.}~\bibnamefont {Blackman}} \emph {et~al.},\ }\href {https://doi.org/10.1103/PhysRevD.96.024058} {\bibfield  {journal} {\bibinfo  {journal} {Physical Review D}\ }\textbf {\bibinfo {volume} {96}},\ \bibinfo {pages} {1} (\bibinfo {year} {2017})},\ \Eprint {https://arxiv.org/abs/1705.07089} {arXiv:1705.07089} \BibitemShut {NoStop}%
\bibitem [{\citenamefont {Rezzolla}\ \emph {et~al.}(2010)\citenamefont {Rezzolla}, \citenamefont {MacEdo},\ and\ \citenamefont {Jaramillo}}]{Rezzolla2010}%
  \BibitemOpen
  \bibfield  {author} {\bibinfo {author} {\bibfnamefont {L.}~\bibnamefont {Rezzolla}}, \bibinfo {author} {\bibfnamefont {R.~P.}\ \bibnamefont {MacEdo}},\ and\ \bibinfo {author} {\bibfnamefont {J.~L.}\ \bibnamefont {Jaramillo}},\ }\href {https://doi.org/10.1103/PhysRevLett.104.221101} {\bibfield  {journal} {\bibinfo  {journal} {Physical Review Letters}\ }\textbf {\bibinfo {volume} {104}},\ \bibinfo {pages} {4} (\bibinfo {year} {2010})},\ \Eprint {https://arxiv.org/abs/1003.0873} {arXiv:1003.0873} \BibitemShut {NoStop}%
\bibitem [{\citenamefont {Owen}\ \emph {et~al.}(2011)\citenamefont {Owen} \emph {et~al.}}]{Owen2011}%
  \BibitemOpen
  \bibfield  {author} {\bibinfo {author} {\bibfnamefont {R.}~\bibnamefont {Owen}} \emph {et~al.},\ }\href {https://doi.org/10.1103/PhysRevLett.106.151101} {\bibfield  {journal} {\bibinfo  {journal} {Physical Review Letters}\ }\textbf {\bibinfo {volume} {106}},\ \bibinfo {pages} {4} (\bibinfo {year} {2011})},\ \Eprint {https://arxiv.org/abs/1012.4869} {arXiv:1012.4869} \BibitemShut {NoStop}%
\bibitem [{\citenamefont {Nichols}\ \emph {et~al.}(2011)\citenamefont {Nichols} \emph {et~al.}}]{Nichols2011a}%
  \BibitemOpen
  \bibfield  {author} {\bibinfo {author} {\bibfnamefont {D.~A.}\ \bibnamefont {Nichols}} \emph {et~al.},\ }\href {https://doi.org/10.1103/PhysRevD.84.124014} {\bibfield  {journal} {\bibinfo  {journal} {Physical Review D - Particles, Fields, Gravitation and Cosmology}\ }\textbf {\bibinfo {volume} {84}},\ \bibinfo {pages} {1} (\bibinfo {year} {2011})},\ \Eprint {https://arxiv.org/abs/1108.5486} {arXiv:1108.5486} \BibitemShut {NoStop}%
\bibitem [{\citenamefont {Zhang}\ \emph {et~al.}(2012)\citenamefont {Zhang} \emph {et~al.}}]{Zhang2012}%
  \BibitemOpen
  \bibfield  {author} {\bibinfo {author} {\bibfnamefont {F.}~\bibnamefont {Zhang}} \emph {et~al.},\ }\href {https://doi.org/10.1103/PhysRevD.86.084049} {\bibfield  {journal} {\bibinfo  {journal} {Physical Review D - Particles, Fields, Gravitation and Cosmology}\ }\textbf {\bibinfo {volume} {86}},\ \bibinfo {pages} {1} (\bibinfo {year} {2012})},\ \Eprint {https://arxiv.org/abs/1208.3034} {arXiv:1208.3034} \BibitemShut {NoStop}%
\bibitem [{\citenamefont {Nichols}\ \emph {et~al.}(2012)\citenamefont {Nichols} \emph {et~al.}}]{Nichols2012b}%
  \BibitemOpen
  \bibfield  {author} {\bibinfo {author} {\bibfnamefont {D.~A.}\ \bibnamefont {Nichols}} \emph {et~al.},\ }\href {https://doi.org/10.1103/PhysRevD.86.104028} {\bibfield  {journal} {\bibinfo  {journal} {Physical Review D - Particles, Fields, Gravitation and Cosmology}\ }\textbf {\bibinfo {volume} {86}},\ \bibinfo {pages} {1} (\bibinfo {year} {2012})},\ \Eprint {https://arxiv.org/abs/1208.3038} {arXiv:1208.3038} \BibitemShut {NoStop}%
\bibitem [{\citenamefont {Jaramillo}\ \emph {et~al.}(2012{\natexlab{a}})\citenamefont {Jaramillo}, \citenamefont {Macedo}, \citenamefont {Moesta},\ and\ \citenamefont {Rezzolla}}]{Jaramillo2012}%
  \BibitemOpen
  \bibfield  {author} {\bibinfo {author} {\bibfnamefont {J.~L.}\ \bibnamefont {Jaramillo}}, \bibinfo {author} {\bibfnamefont {R.~P.}\ \bibnamefont {Macedo}}, \bibinfo {author} {\bibfnamefont {P.}~\bibnamefont {Moesta}},\ and\ \bibinfo {author} {\bibfnamefont {L.}~\bibnamefont {Rezzolla}},\ }\href {https://doi.org/10.1063/1.4734411} {\bibfield  {journal} {\bibinfo  {journal} {AIP Conference Proceedings}\ }\textbf {\bibinfo {volume} {1458}},\ \bibinfo {pages} {158} (\bibinfo {year} {2012}{\natexlab{a}})}\BibitemShut {NoStop}%
\bibitem [{\citenamefont {Jaramillo}\ \emph {et~al.}(2012{\natexlab{b}})\citenamefont {Jaramillo}, \citenamefont {MacEdo}, \citenamefont {Moesta},\ and\ \citenamefont {Rezzolla}}]{Jaramillo2012a}%
  \BibitemOpen
  \bibfield  {author} {\bibinfo {author} {\bibfnamefont {J.~L.}\ \bibnamefont {Jaramillo}}, \bibinfo {author} {\bibfnamefont {R.~P.}\ \bibnamefont {MacEdo}}, \bibinfo {author} {\bibfnamefont {P.}~\bibnamefont {Moesta}},\ and\ \bibinfo {author} {\bibfnamefont {L.}~\bibnamefont {Rezzolla}},\ }\href {https://doi.org/10.1103/PhysRevD.85.084030} {\bibfield  {journal} {\bibinfo  {journal} {Physical Review D - Particles, Fields, Gravitation and Cosmology}\ }\textbf {\bibinfo {volume} {85}},\ \bibinfo {pages} {1} (\bibinfo {year} {2012}{\natexlab{b}})},\ \Eprint {https://arxiv.org/abs/1108.0060} {arXiv:1108.0060} \BibitemShut {NoStop}%
\bibitem [{\citenamefont {Jaramillo}\ \emph {et~al.}(2012{\natexlab{c}})\citenamefont {Jaramillo}, \citenamefont {MacEdo}, \citenamefont {Moesta},\ and\ \citenamefont {Rezzolla}}]{Jaramillo2012b}%
  \BibitemOpen
  \bibfield  {author} {\bibinfo {author} {\bibfnamefont {J.~L.}\ \bibnamefont {Jaramillo}}, \bibinfo {author} {\bibfnamefont {R.~P.}\ \bibnamefont {MacEdo}}, \bibinfo {author} {\bibfnamefont {P.}~\bibnamefont {Moesta}},\ and\ \bibinfo {author} {\bibfnamefont {L.}~\bibnamefont {Rezzolla}},\ }\href {https://doi.org/10.1103/PhysRevD.85.084031} {\bibfield  {journal} {\bibinfo  {journal} {Physical Review D - Particles, Fields, Gravitation and Cosmology}\ }\textbf {\bibinfo {volume} {85}},\ \bibinfo {pages} {1} (\bibinfo {year} {2012}{\natexlab{c}})},\ \Eprint {https://arxiv.org/abs/1108.0061} {arXiv:1108.0061} \BibitemShut {NoStop}%
\bibitem [{\citenamefont {Gupta}\ \emph {et~al.}(2018)\citenamefont {Gupta}, \citenamefont {Krishnan}, \citenamefont {Nielsen},\ and\ \citenamefont {Schnetter}}]{Gupta2018}%
  \BibitemOpen
  \bibfield  {author} {\bibinfo {author} {\bibfnamefont {A.}~\bibnamefont {Gupta}}, \bibinfo {author} {\bibfnamefont {B.}~\bibnamefont {Krishnan}}, \bibinfo {author} {\bibfnamefont {A.~B.}\ \bibnamefont {Nielsen}},\ and\ \bibinfo {author} {\bibfnamefont {E.}~\bibnamefont {Schnetter}},\ }\href {https://doi.org/10.1103/PhysRevD.97.084028} {\bibfield  {journal} {\bibinfo  {journal} {Physical Review D}\ }\textbf {\bibinfo {volume} {97}},\ \bibinfo {pages} {84028} (\bibinfo {year} {2018})},\ \Eprint {https://arxiv.org/abs/1801.07048} {arXiv:1801.07048} \BibitemShut {NoStop}%
\bibitem [{\citenamefont {Prasad}\ \emph {et~al.}(2020)\citenamefont {Prasad} \emph {et~al.}}]{Prasad2020}%
  \BibitemOpen
  \bibfield  {author} {\bibinfo {author} {\bibfnamefont {V.}~\bibnamefont {Prasad}} \emph {et~al.},\ }\href {https://doi.org/10.1103/PhysRevLett.125.121101} {\bibfield  {journal} {\bibinfo  {journal} {Physical Review Letters}\ }\textbf {\bibinfo {volume} {125}},\ \bibinfo {pages} {121101} (\bibinfo {year} {2020})},\ \Eprint {https://arxiv.org/abs/2003.06215} {arXiv:2003.06215} \BibitemShut {NoStop}%
\bibitem [{\citenamefont {Mourier}\ \emph {et~al.}(2021)\citenamefont {Mourier}, \citenamefont {{Jim{\'{e}}nez Forteza}}, \citenamefont {Pook-Kolb}, \citenamefont {Krishnan},\ and\ \citenamefont {Schnetter}}]{Mourier2021}%
  \BibitemOpen
  \bibfield  {author} {\bibinfo {author} {\bibfnamefont {P.}~\bibnamefont {Mourier}}, \bibinfo {author} {\bibfnamefont {X.}~\bibnamefont {{Jim{\'{e}}nez Forteza}}}, \bibinfo {author} {\bibfnamefont {D.}~\bibnamefont {Pook-Kolb}}, \bibinfo {author} {\bibfnamefont {B.}~\bibnamefont {Krishnan}},\ and\ \bibinfo {author} {\bibfnamefont {E.}~\bibnamefont {Schnetter}},\ }\href {https://doi.org/10.1103/PhysRevD.103.044054} {\bibfield  {journal} {\bibinfo  {journal} {Physical Review D}\ }\textbf {\bibinfo {volume} {103}},\ \bibinfo {pages} {1} (\bibinfo {year} {2021})},\ \Eprint {https://arxiv.org/abs/2010.15186} {arXiv:2010.15186} \BibitemShut {NoStop}%
\bibitem [{\citenamefont {Prasad}\ \emph {et~al.}(2022)\citenamefont {Prasad}, \citenamefont {Gupta}, \citenamefont {Bose},\ and\ \citenamefont {Krishnan}}]{Prasad2022}%
  \BibitemOpen
  \bibfield  {author} {\bibinfo {author} {\bibfnamefont {V.}~\bibnamefont {Prasad}}, \bibinfo {author} {\bibfnamefont {A.}~\bibnamefont {Gupta}}, \bibinfo {author} {\bibfnamefont {S.}~\bibnamefont {Bose}},\ and\ \bibinfo {author} {\bibfnamefont {B.}~\bibnamefont {Krishnan}},\ }\href {https://doi.org/10.1103/PhysRevD.105.044019} {\bibfield  {journal} {\bibinfo  {journal} {Physical Review D}\ }\textbf {\bibinfo {volume} {105}},\ \bibinfo {pages} {44019} (\bibinfo {year} {2022})},\ \Eprint {https://arxiv.org/abs/2106.02595} {arXiv:2106.02595} \BibitemShut {NoStop}%
\bibitem [{\citenamefont {Khera}\ \emph {et~al.}(2023)\citenamefont {Khera} \emph {et~al.}}]{Khera2023}%
  \BibitemOpen
  \bibfield  {author} {\bibinfo {author} {\bibfnamefont {N.}~\bibnamefont {Khera}} \emph {et~al.},\ }\href {https://doi.org/10.1103/PhysRevLett.131.231401} {\bibfield  {journal} {\bibinfo  {journal} {Physical Review Letters}\ }\textbf {\bibinfo {volume} {131}},\ \bibinfo {pages} {1} (\bibinfo {year} {2023})},\ \Eprint {https://arxiv.org/abs/2306.11142} {arXiv:2306.11142} \BibitemShut {NoStop}%
\bibitem [{\citenamefont {Prasad}(2023)}]{Prasad2023}%
  \BibitemOpen
  \bibfield  {author} {\bibinfo {author} {\bibfnamefont {V.}~\bibnamefont {Prasad}},\ }\href {https://doi.org/10.1103/PhysRevD.111.084070} {\bibfield  {journal} {\bibinfo  {journal} {Physical Review D}\ }\textbf {\bibinfo {volume} {111}},\ \bibinfo {pages} {84070} (\bibinfo {year} {2023})},\ \Eprint {https://arxiv.org/abs/2312.01136} {arXiv:2312.01136} \BibitemShut {NoStop}%
\bibitem [{\citenamefont {Lange}\ \emph {et~al.}(2018)\citenamefont {Lange}, \citenamefont {O'Shaughnessy},\ and\ \citenamefont {Rizzo}}]{lange2018rapid}%
  \BibitemOpen
  \bibfield  {author} {\bibinfo {author} {\bibfnamefont {J.}~\bibnamefont {Lange}}, \bibinfo {author} {\bibfnamefont {R.}~\bibnamefont {O'Shaughnessy}},\ and\ \bibinfo {author} {\bibfnamefont {M.}~\bibnamefont {Rizzo}},\ }\Eprint {https://arxiv.org/abs/1805.10457} {arXiv:1805.10457 [gr-qc]}  (\bibinfo {year} {2018})\BibitemShut {NoStop}%
\bibitem [{\citenamefont {Ashton}\ \emph {et~al.}(2019)\citenamefont {Ashton} \emph {et~al.}}]{Ashton_2019}%
  \BibitemOpen
  \bibfield  {author} {\bibinfo {author} {\bibfnamefont {G.}~\bibnamefont {Ashton}} \emph {et~al.},\ }\href {https://doi.org/10.3847/1538-4365/ab06fc} {\bibfield  {journal} {\bibinfo  {journal} {The Astrophysical Journal Supplement Series}\ }\textbf {\bibinfo {volume} {241}},\ \bibinfo {pages} {27} (\bibinfo {year} {2019})}\BibitemShut {NoStop}%
\bibitem [{\citenamefont {Veitch}\ \emph {et~al.}(2015)\citenamefont {Veitch} \emph {et~al.}}]{Veitch_2015}%
  \BibitemOpen
  \bibfield  {author} {\bibinfo {author} {\bibfnamefont {J.}~\bibnamefont {Veitch}} \emph {et~al.},\ }\bibfield  {journal} {\bibinfo  {journal} {Physical Review D}\ }\textbf {\bibinfo {volume} {91}},\ \href {https://doi.org/10.1103/physrevd.91.042003} {10.1103/physrevd.91.042003} (\bibinfo {year} {2015})\BibitemShut {NoStop}%
\bibitem [{\citenamefont {Virtanen}\ \emph {et~al.}(2020)\citenamefont {Virtanen} \emph {et~al.}}]{SciPy}%
  \BibitemOpen
  \bibfield  {author} {\bibinfo {author} {\bibfnamefont {P.}~\bibnamefont {Virtanen}} \emph {et~al.},\ }\href {https://doi.org/10.1038/s41592-019-0686-2} {\bibfield  {journal} {\bibinfo  {journal} {Nature Methods}\ }\textbf {\bibinfo {volume} {17}},\ \bibinfo {pages} {261} (\bibinfo {year} {2020})}\BibitemShut {NoStop}%
\end{thebibliography}%
\end{document}